\documentclass[twocolumn,prb,tightenlines,superscriptaddress,amsmath,amssymb,amsfonts]{revtex4-2}
\usepackage{graphicx}
\usepackage{dcolumn}
\usepackage{bm}
\usepackage{float}
\usepackage{hyperref}
\usepackage[section]{placeins}
\usepackage{epstopdf}
\usepackage[table, dvipsnames]{xcolor}
\usepackage[T1]{fontenc}
\usepackage{dcolumn}

\DeclareMathOperator{\sinc}{sinc}

\newcommand{\ie}{\textit{i}.\textit{e}.}

\begin{document}

\title{Quasiparticle states for integer- and fractional-charged electron wave packets}
\author{X. K. Yue}
\author{Y. Yin}
\thanks{Author to  whom correspondence should be addressed}
\email{yin80@scu.edu.cn.}
\affiliation{Department of Physics,
  Sichuan University, Chengdu, Sichuan, 610065, China}
\date{\today}

\begin{abstract}
  It is well-known that Lorentzian voltage pulses with integer quantum flux can
  inject integer-charged wave packets without electron-hole pairs. The wave
  packets are composed of soliton-like quasiparticles on top of the Fermi sea,
  which have been named as ``levitons''. However, it is not clear what kind of
  charged quasiparticles can be injected by Lorentzian pulses with fractional
  quantum flux. To answer this question, we study the wave packets injected by a
  train of Lorentzian pulses with repetition period $T$. We introduce a set of
  single-body wave functions, within which the quantum state of the charged
  quasiparticles can be described for pulses with arbitrary quantum flux. We
  find that in general case the charged quasiparticles are injected with two
  different periods. For pulses with integer quantum flux, the two periods match
  each other. The charged quasiparticles are levitons in this case, which are
  injected with a single period $T$. For pulses with fractional quantum flux,
  the two periods mismatch. The charged quasiparticles can then be injected in a
  multi-periodic way. This makes the each quasiparticle can carry only a
  fractional electric charge into the quantum conductor within a single period
  $T$. These quasiparticles can have pronounced impact on the charge
  injection. In particular, they can lead to the cycle-missing event, in which
  the voltage pulse fails to inject an electron within a single period $T$. The
  cycle-missing event can be seen intuitively from the waiting time distribution
  between electrons above the Fermi sea, which exhibits a series of peaks at
  multiplies of the period $T$. By using the wave functions of the charged
  quasiparticles, we elucidate in detail how a leviton evolves as the flux of
  the pulse changes. In the meantime, we also clarify how additional $e$h pairs
  can be excited.
\end{abstract}


\maketitle

\section{Introduction}
\label{sec1}

\begin{figure}[H]
  \includegraphics[width=7.0cm]{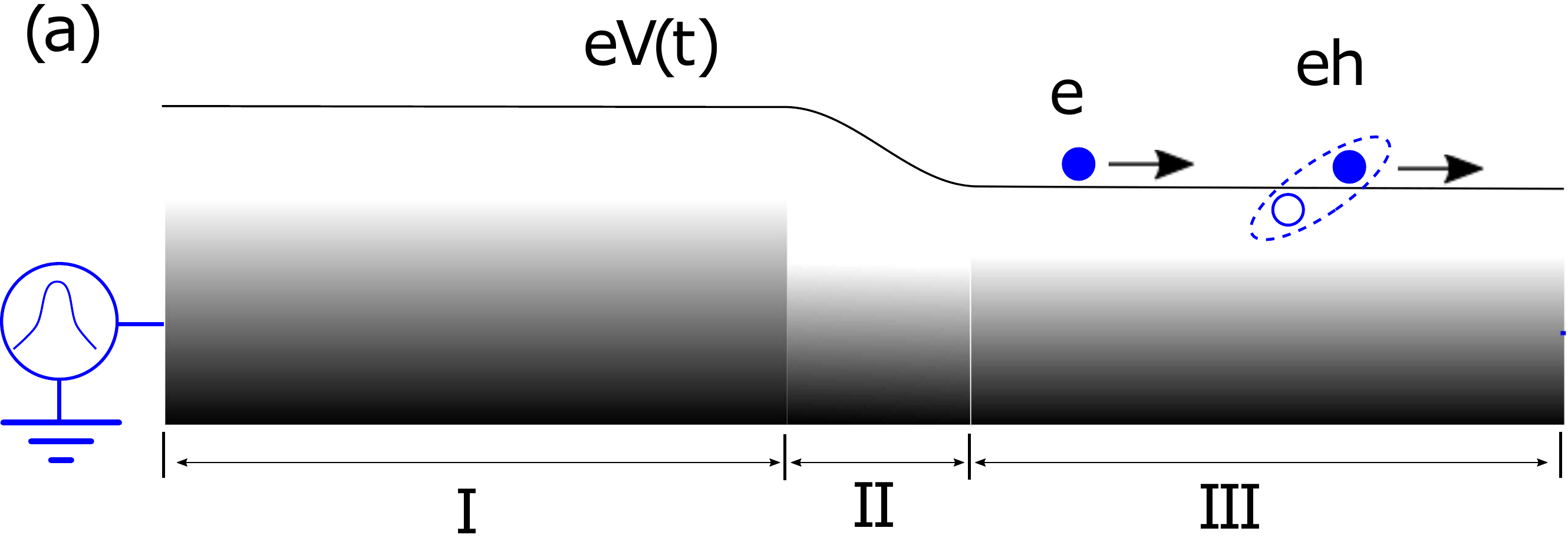}
  \includegraphics[width=7.0cm]{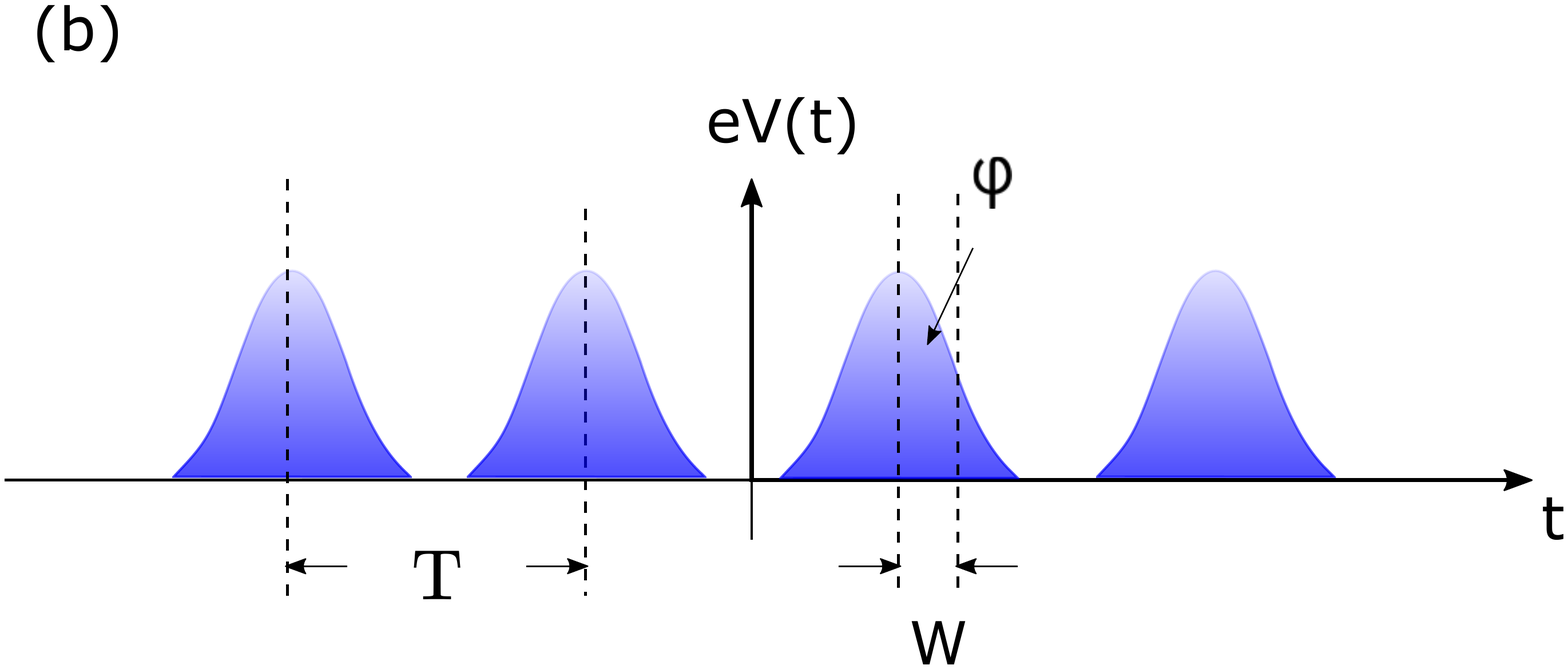}
  \caption{ (a) Schematic of the on-demand electron injection via the voltage
    pulse $V(t)$. By applying $V(t)$ on the contact of the quantum conductor,
    electron (hole) or $e$h pairs from the reservoir (region I) can be injected
    into the quantum conductor (region III). The voltage drop is assumed to
    occur across a short interval at the interface (region II). (b) Schematic of
    the applied voltage pulse train. The pulse train is composed of identical
    Lorentzian pulses, which can be characterized by half width at half maximum
    $W$ and Faraday flux $\varphi$. These pulses are separated by a time
    interval $T$.}
  \label{fig1}
\end{figure}

In the past decade, much effort has been devoted to the on-demand
single-electron source, within which electron wave packets carrying single or
few electric charges can be injected coherently into a quantum conductor
\cite{dubois-2013-minim-excit, dubois-2013-integ-fract,
  keeling-2006-minim-excit, bocquillon-2013-coher-indis, hofer-2013-emiss-time,
  moskalets-2013-singl-elect-sourc, gabelli-2013-shapin-time,
  jullien-2014-quant-tomog-elect, feve-2007-deman-coher,
  keeling-2008-coher-partic, mahe-2010-curren-correl, albert-2010-accur-quant,
  grenier-2011-singl-elect, haack-2011-coher-singl, sherkunov-2012-optim-pumpin,
  bocquillon-2012-elect-quant-optic, fletcher-2013-clock-contr,
  ubbelohde-2014-partit-deman, ryu-2016-ultraf-emiss,
  splettstoesser-2017-singl-elect, misiorny-2018-shapin-charg,
  dashti-2019-minim-excit}. In a simple way, such injection can be realized by
applying a nanosecond pulse on the Ohmic contact of the conductor, as
illustrated in Fig.~\ref{fig1}. The injected charges $Q$ of the wave packet is
decided by the flux $\varphi$ of the pulse, while the detailed quantum states of
the wave packet can be controlled via fine-tuning the profile of the pulse. This
offers a simple but feasible approach to archive the time-resolved quantum
control of propagating electron wave packet in solid-state circuits
\cite{grenier-2013-fract-minim, wahl-2014-inter-charg, ferraro-2014-real-time,
  kamata-2014-fract-wave, freulon-2015-hong-ou, dasenbrook-2015-dynam-gener,
  belzig-2016-elemen-andreev, vannucci-2017-minim-excit, rech-2017-minim-excit,
  yin-2018-deman-elect, ronetti-2018-cryst-levit, vanevic-2016-elect-elect,
  vanevic-2012-contr-elect, bisognin-2019-quant-tomog}.

Generally speaking, the wave packet is composed of charged quasiparticles in the
Fermi sea ($| \mathbf{F} \rangle$) of the conductor, which are usually
accompanied by a neutral cloud of electron-hole ($e$h) pairs
\cite{landau-1957-theor-fermi-liquid,
  pines-2018-theor-quant-liquid}. Remarkably, it is possible to inject a
``clean'' wave packet without $e$h pairs, which can be done by tuning the pulse
to be a Lorentzian with integer quantum flux \cite{ivanov-1997-coher-states,
  keeling-2006-minim-excit}. In doing so, one obtains soliton-like
quasiparticles propagating on top of the Fermi sea, which has been named as
``levitons'' \cite{dubois-2013-integ-fract,
  jullien-2014-quant-tomog-elect}. Each leviton carries an unit electric charge
and has a well-defined wave function. A sequence of well-separated levitons can
be injected by using a train of Lorentzian pulses, emerging as promising
candidates for flying qubits in solid state circuits
\cite{bocquillon-2014-elect-quant, dasenbrook-2016-singl-elect,
  glattli-2016-levit-elect, baeuerle-2018-coher-contr,
  olofsson-2020-quant-telep}.

By using a Lorentzian pulse with fractional quantum flux, one can inject a wave
packet carrying fractional charges, which has a quite different structure. On
one hand, it contains a large amount of $e$h pairs, which is closely related to
the dynamical orthogonality catastrophe \cite{levitov-1996-elect-count,
  glattli-2018-pseud-binar}. On the other hand, it can sustain charged
quasiparticles carrying fractional charges. The structure of the wave packet has
been demonstrated for the Lorentzian pulse with a half-quantum flux. In this
case, the quantum state of the wave packet can be decomposed into two mixed
states: one represents the neutral cloud of $e$h pairs, while the other one can
be regarded as a zero-energy quasiparticle carrying an effective $e/2$ charge
\cite{moskalets-2016-fract-charg}. This makes the wave packet show distinctly
different features from the wave packet built from levitons
\cite{hofer-2014-mach-zehnd, gaury-2014-dynam-contr,
  belzig-2016-elemen-andreev}.

Intuitively, one expect that the fractional-charged quasiparticles can be
injected in a similar way as levitons, providing an alternative approach to
realize flying qubits. However, the nature of these quasiparticles has not been
fully understood yet. In particular, it is not clear how a leviton can evolve
into a fractional-charged quasiparticle as the flux of pulses changes. To answer
this question, one needs to describe the quantum states for both integer- and
fractional-charged wave packets in an unified manner, which has not been given
yet.

In this paper, we attack this problem by examining the case when a Lorentzian
pulse train with repetition period $T$ is applied on the Ohmic contact, as
illustrated in Fig.~\ref{fig1}(b). In this case, we show that the injected
charges are carried by a train of of wave packets, whose quantum state can be
given as
\begin{equation}
  | \mathbf{\Psi_{\rm train}} \rangle = \prod_{l=0, \pm 1, \pm 2, ...} | \mathbf{\Psi}_l \rangle,
  \label{s1:eq1}
\end{equation}
with $| \mathbf{\Psi}_l \rangle$ representing the quantum state of the $l$-th
wave packet. Each wave packet is composed of charged quasiparticles and neutral
$e$h pairs, which can be described by a set of single-body wave functions
$\psi^\alpha_{kl}(t)$, with $\alpha=c$ for the quasiparticles and $\alpha=e$/$h$
for the electron/hole component of the $e$h pairs. This allows one to introduce
the corresponding creation operators as
\begin{align}
  C^{\dagger}_{kl} & = \int^{+\infty}_{-\infty} dt \psi^c_{kl}(t) \hat{a}^{\dagger}(t), \nonumber\\
  (B^e_{kl})^{\dagger} & = \int^{+\infty}_{-\infty} dt \psi^e_{kl}(t) \hat{a}^{\dagger}(t), \nonumber\\
  (B^h_{kl})^{\dagger} & = \int^{+\infty}_{-\infty} dt \psi^h_{kl}(t) \hat{a}(t),
                         \label{s1:eq2}
\end{align}
with $\hat{a}(t)$ [$\hat{a}^{\dagger}(t)$] being the electron annihilation
[creation] operator in the time domain. In doing so, the quantum state of the
$l$-th wave packet can be described by the Slater determinant as
\begin{equation}
  | \mathbf{\Psi}_l \rangle = \Big[\prod_k C^{\dagger}_{kl}\Big] \prod_k \Big[\sqrt{1 - p_k} + i
  \sqrt{p_k}(B^e_{kl})^{\dagger} (B^h_{kl})^{\dagger} \Big] | \mathbf{F}\rangle,
  \label{s1:eq3}
\end{equation}
with $p_k$ representing the excitation probabilities of the $e$h pairs. Both the
excitation probabilities $p_k$ and the single-body wave functions
$\psi^\alpha_{kl}(t)$ can be extracted from the time-dependent scattering
matrix, providing a general way to study the quantum state of both the integer-
and fractional-charged wave packets.

\begin{figure}
  \includegraphics[width=8.0cm]{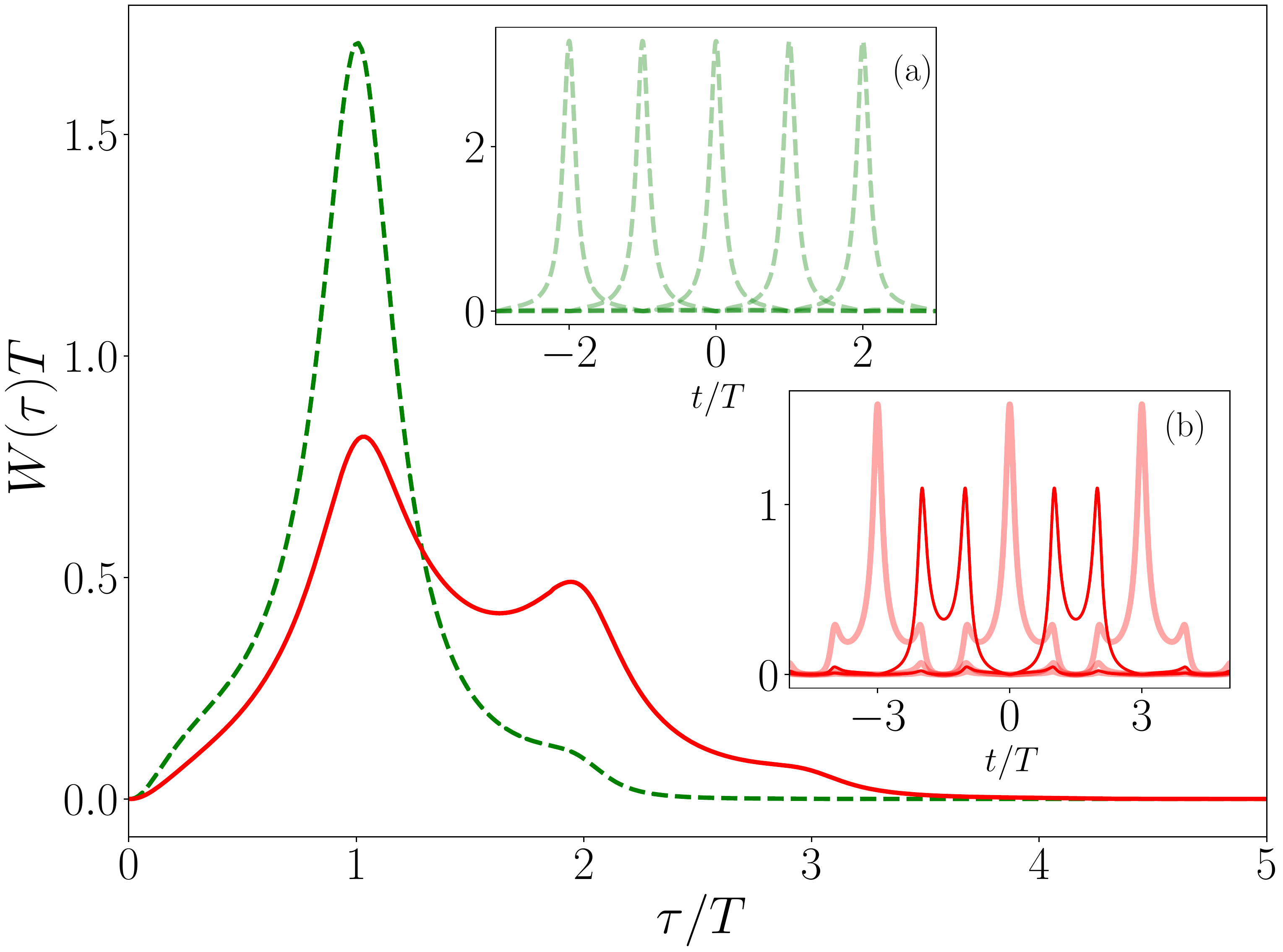}
  \caption{ (Color online) The waiting time distribution $W(\tau)$ between
    electrons above the Fermi sea (main panel) and the corresponding wave
    function $\psi^c_{kl}(t)$ (inset), corresponding to the pulse width
    $W/T=0.1$. The green dashed curve in the inset (a) represent the wave
    functions of levitons, corresponding to $Q/e=1$. The red solid curves in the
    inset (b) represent the wave functions of the charged quasiparticles,
    corresponding to $Q/e=2/3$. Note that there are two types of quasiparticles
    here, which are represented by the thick and thin curves.}
  \label{fig2}
\end{figure}

As the charges $Q$ are injected with the period $T$, one may expect that the
charged quasiparticles are also injected with the same period. Indeed, this
picture holds when $Q/e$ takes integer values. This is illustrated in the inset
of Fig.~\ref{fig2}, corresponding to $Q/e=1$. In this case, all the single-body
wave functions of the charged quasiparticles exhibit the same profiles, which
are separated from each other by the time interval $T$. They essentially
correspond to a periodic train of levitons. The structure of the leviton train
can be understood intuitively from the corresponding waiting time distribution
$W(\tau)$ (WTD) \cite{dasenbrook16_quant_theor_elect_waitin_time_clock}, which
is characterized by a strong peak around $\tau=T$ [see the green dashed curve in
the main panel of Fig.~\ref{fig2}]. This indicates that the voltage pulse tends
to inject exactly one electron per period into the quantum conductor.

In contrast, the above picture is inapplicable when $Q/e$ takes fractional
values. In this case, the charged quasiparticles are essentially injected with
two different periods. Due to the mismatch between these two periods, the wave
functions of the quasiparticles can exhibit different profiles, which are
injected with an extended period longer than $T$. This is illustrated in the
inset (b) of Fig.~\ref{fig2}, corresponding to $Q/e=2/3$. One can see that the
wave functions can exhibit two types of profiles, which are plotted with thick
and thin curves. They are separated from each other by the time interval
$3T/2$. On average, each quasiparticle can carry only $2e/3$ charge into the
quantum conductor within a single period $T$. This makes them behave {\em
  effectively} like quasiparticles carrying fractional charges. These
quasiparticles can have pronounced impact on the charge injection. In
particular, they lead to the cycle-missing event, in which the voltage pulse can
fail to inject an electron within a single period $T$. Such event can be seen
from the corresponding WTD $W(\tau)$, which exhibits a series of peaks at
multiplies of the period $T$ [see the red solid curve in Fig.~\ref{fig2}].

The wave function $\psi^c_{kl}(t)$ can hence provide an unified description of
the charged quasiparticles, which is applicable for both the integer- and
fractional-charged wave packets. This allows us to elucidate in detail how a
leviton can evolve into a fractional-charged quasiparticle as the flux $\varphi$
of the pulse changes. In the meantime, our approach can also provide the
information of the $e$h pairs. This allow us to clarify how additional $e$h
pairs can be excited during the evolution of the levitons.

The paper is organized as follows: In Sec.~\ref{sec2}, we present the model of
the system and introduce a general expression for the quantum state of the wave
packets. We discuss the typical behaviors of the wave functions of
quasiparticles in Sec.~\ref{sec3} and~\ref{sec4}. The corresponding waiting time
distribution is also discussed in these two sections. The evolution of levitons
and $e$h pairs are discussed in Sec.~\ref{sec5} and~\ref{sec6}, respectively. We
summarizes in Sec.~\ref{sec7}.

\section{Bloch-Messiah reduction in the framework of scattering matrix formalism}
\label{sec2}

The electron source can be modeled as a single-mode quantum conductor, as
illustrated in Fig.~\ref{fig1}(a). We choose the driving voltage $V(t)$ of the
form
\begin{equation}
  \frac{e}{\hbar}V(t) =  \sum_{l=0, \pm 1, \pm 2, ...}  \frac{2 \varphi W}{W^2 + (t-lT)^2},
  \label{s2:eq0}
\end{equation}
which corresponds to a periodic train of Lorentzian pulses with width $W$ [see
Fig.~\ref{fig1}(b)]. The voltage drop $V(t)$ between the contact and the
conductor is assumed to occur across a short interval, so that the corresponding
dwell time $\tau_D$ satisfies:
$ k_B T_e \ll \hbar/T < \hbar/W \ll \hbar/\tau_D \ll E_F$, with $E_F$
representing the Fermi energy and $T_e$ representing the electron
temperature. In this paper, we choose $E_F=0$ and concentrate on the
zero-temperature limit.

The scattering matrix of the system can be solely determined by the driving
voltage $V(t)$ as
\begin{equation}
  S(t, t') = \delta(t-t') \exp[- i \frac{e}{\hbar} \int^t_0 d\tau V(\tau)].
  \label{s2:eq1}
\end{equation}
Given the scattering matrix, the electrons in the contact and the conductor can
be related via the equation
\begin{eqnarray}
  \hat{b}(t) = \int dt' S(t, t') \hat{a}(t'),
  \label{s2:eq2}
\end{eqnarray}
where $\hat{a}(t)$ and $\hat{b}(t)$ represent the electron annihilation
operators in the Ohmic contact and the quantum conductor, respectively.

In this setup, the injected current can be simply given as
$I(t) = (e^2/h) V(t)$. The charge $Q$ injected within a single period can be
solely by the flux $\varphi$ as
\begin{eqnarray}
  Q = \int^{+T/2}_{-T/2} dt I(t) = e \varphi.
  \label{s2:eq3}
\end{eqnarray}
For simplicity, here we assume $Q/e>0$ so that the wave packets carry negative
charges.

The quantum state of the injected wave packets can be obtained from the
Bloch-Messiah reduction, which extracts the many-body quantum state from the
decomposition of the first-order correlation function $G(t, t')$
\cite{yin-2019-quasip-states, yue-2019-normal-anomal}. In the zero-temperature
limit, $G(t, t')$ can be given as
\begin{equation}
  i G(t, t') = \langle \mathbf{F} | \hat{b}^{\dagger}(t') \hat{b}(t) | \mathbf{F} \rangle,
  \label{s2:eq4}
\end{equation}
with $| \mathbf{F} \rangle$ representing the Fermi sea. To find the many-body
state corresponding to $G(t, t')$, the Bloch-Messiah reduction essentially seeks
out the quantum state $| \mathbf{\Psi} \rangle$, which satisfies
\begin{equation}
  \langle \mathbf{\Psi} | \hat{a}^{\dagger}(t') \hat{a}(t) | \mathbf{\Psi} \rangle = \langle \mathbf{F} | \hat{b}^{\dagger}(t') \hat{b}(t) | \mathbf{F} \rangle.
  \label{s2:eq5}
\end{equation}
This can be done by a proper decomposition of $G(t, t')$. Here we only present
the outline and leave the details to Appendix~\ref{app1}.

\subsection{Decomposition in Floquet space}

For the system under periodic driving, it is straightforward to perform such decomposition in Floquet space
\cite{moskalets-2011-scatt-matrix, pedersen-1998-scatt-theor}, which can be generally written as
\begin{align}
  \label{s2:eq6}  
  i G(t, t') &= \sum_k \int^{\Omega}_0 \frac{d\omega}{\Omega} e^{-i\omega(t-t')}
               u^c_k(\omega, t) [ u^c_k(\omega, t') ]^{\ast} \\
             & \mbox{}+ \sum_k \int^{\Omega}_0 \frac{d\omega}{\Omega} e^{-i\omega(t-t')} \left[ \begin{tabular}{cc}
                                                                                                 $u^e_k(\omega, t)$, & $u^h_k(\omega, t)$\\
                                                                                               \end{tabular}\right] \nonumber\\
             & \hspace{-1.5cm}\times \left[\begin{tabular}{cc}
                                             $p_k(\omega)$ & $i \sqrt{p_k(\omega)[1 - p_k(\omega)]}$\\
                                             $-i \sqrt{p_k(\omega)[1 - p_k(\omega)]}$ & $1 - p_k(\omega)$\\
                                           \end{tabular}\right] \left[\begin{tabular}{c}
                                                                        $u^e_k(\omega, t')$\\
                                                                        $u^h_k(\omega, t')$\\
                                                                      \end{tabular}\right]^{\ast}, \nonumber
\end{align}
with asterisk denoting the complex conjugation. In the above expression, the quantity $p_k(\omega)$ is real, which
satisfies $p_k(\omega) \in [0, 1]$. The functions $u^\alpha_k(\omega, t)$ are complex, which are periodic in the time
domain $u^\alpha_k(\omega, t) = u^\alpha_k(\omega, t+T)$ with $\alpha = c, e$ and $h$. These functions can form
orthonormal basis within a single period, \ie,
\begin{equation}
  \label{s2:eq7}
  \int^{T/2}_{-T/2} dt [ u^{\alpha'}_{k'}(\omega, t) ]^\ast  u^\alpha_k(\omega,
  t) = \delta_{\alpha, \alpha'} \delta_{k, k'}.
\end{equation}
All these functions can be characterized by two indices $\omega$ and $k$. Here $k$ is a discrete index, which can be
described by (dimensionless) integer numbers. In contrast, the index $\omega$ has the unit of frequency, which satisfies
$\omega \in [0, \Omega)$ with $\Omega = 2\pi/T$ being the repetition rate of the pulses.

The function $u^\alpha_k(\omega, t)$ is closely related to the single-body wave function of the charged quasiparticle
($\alpha=c$) and the neutral $e$h pair ($\alpha=e, h$), while $p_k(\omega)$ represents the excitation probability of the
$e$h pair. Both $u^\alpha_k(\omega, t)$ and $p_k(\omega)$ can be obtained from the polar decomposition of the scattering
matrix. In general cases, they can exhibit a complicated dependence on $\omega$. For the scattering matrix given in
Eq.~\eqref{s2:eq1}, we find that the $\omega$-dependence can be much simpler: First, the probabilities $p_k(\omega)$ are
independent on $\omega$ and can hence be written as $p_k$ for short. Second, $u^\alpha_k(\omega, t)$ can be written in
the form of separation of variables as
\begin{equation}
  u^\alpha_k(\omega, t) = U^\alpha_k(t)F^{Q}_k(\omega),
  \label{s2:eq8}
\end{equation}
where $F^{Q}_k(\omega)$ is a real function defined in the region $\omega \in [0, \Omega)$, while $U^\alpha_k(t)$ is a
complex periodic function defined in the whole time domain $t \in (-\infty, +\infty)$, which satisfies
$U^\alpha_k(t) = U^\alpha_k(t+T)$.

The function $U^\alpha_k(t)$ usually has to be obtained numerically, which is sensitive to the details of the scattering
matrix. In contrast, the function $F^{Q}_k(\omega)$ can be given analytically. To do this, it is convenient to describe
the discrete index $k$ by two non-negative integers $n$ and $m$ [\ie, $n, m = 0, 1, 2, ...$]. In doing so, we find that
$F^{Q}_k(\omega)$ can be written as
\begin{eqnarray}
  F^{Q}_k(\omega) & = & \begin{cases}
    H[ (Q/e - n + 1)\Omega - \omega], & \text{for $Q/e \in [n-1, n]$,}\\
    H[\omega - (Q/e - n)\Omega], & \text{for $Q/e \in (n, n+1]$,}\\
    0, & \text{otherwise.}
  \end{cases}\nonumber\\
  \label{s2:eq9}         
\end{eqnarray}
with $H(\omega)$ representing Heaviside step function \footnote{Here we choose $H(0)=1$.}. Note that $F^{Q}_k(\omega)$
is independent on the details of the scattering matrix and is solely decided by the charged $Q$ of the wave packet.

\begin{table}[h]
  \caption{\label{tab:ex} Parameter space for $k=[n, m]$ for the charged
    quasiparticles ($\alpha=c$) and $e$h pairs ($\alpha=e, h$), corresponding to
    $n, m \le 3$. The parameters for the charged quasiparticles are marked in
    gray shadow.}
  \begin{ruledtabular}  
    \begin{tabular}{|p{2cm}|p{2cm}|p{2cm}|p{2cm}|}
      $[0, 0]$ & \cellcolor{lightgray}$[1, 0]$ & \cellcolor{lightgray}$[2, 0]$
      & \cellcolor{lightgray}$[3, 0]$ \\ \hline
      $[0, 1]$ & $[1, 1]$ & \cellcolor{lightgray}$[2, 1]$ & \cellcolor{lightgray}$[3,1]$ \\ \hline
      $[0, 2]$ & $[1, 2]$ & $[2, 2]$ &\cellcolor{lightgray}$[3, 2]$ \\ \hline
      $[0, 3]$ & $[1, 3]$ & $[2, 3]$ &$[3, 3]$ \\
    \end{tabular}
  \end{ruledtabular}    
\end{table}

It is worth noting that the available parameter space of the index $k=[n, m]$ is different for the charged
quasiparticles ($\alpha=c$) and the $e$h pairs ($\alpha=e, h$): one has $m < n$ for the charged quasiparticles, while
$m \ge n$ for the $e$h pairs. This can be demonstrated more intuitively in Table~\ref{tab:ex}.

\subsection{Decomposition in wave-packet representation}

Given the decomposition of $G(t, t')$ in Eq.~\eqref{s2:eq6}, one can construct a set of single-body wave functions
corresponding to the injected quasiparticles. The many-body state of the wave packets can then be described by using the
Slater determinant built from them. However, one can construct different sets of single-body wave functions, which are
related to each other via unitary transformations. Hence the detailed expression of the Slater determinant is not
uniquely defined. As the driving voltage $V(t)$ corresponds to a train of pulses [see Eq.~\eqref{s2:eq0}], it is
favorable to express the single-body wave functions in a similar form. This can be done by defining a set of wave
functions $\psi^\alpha_{kl}(t)$ from $u^\alpha_k(\omega, t)$ as
\begin{eqnarray}
  \hspace{-0.5cm}\psi^\alpha_{kl}(t) & = & \frac{1}{\sqrt{q_k}}\int^\Omega_0 \frac{d\omega}{\Omega} e^{-i \omega (t-lT/q_k)}
                                           u^\alpha_k(\omega, t), \nonumber\\
                                     & = & U^\alpha_k(t)  \int^\Omega_0 \frac{d\omega}{\Omega}
                                           \frac{F^{Q}_k(\omega)}{\sqrt{q_k}} e^{-i \omega (t-lT/q_k)},
                                           \label{s2:eq10}
\end{eqnarray}
with $l=0, \pm 1, \pm 2, ...$. Note that we have introduced a normalization factor $q_k$ so that $\psi^\alpha_{kl}(t)$
can form an orthonormal basis set in the whole time domain $t \in (-\infty, +\infty)$, which satisfies
\begin{equation}
  \int^{+\infty}_{-\infty}dt  [\psi^{\alpha'}_{k'l'}(t)]^{\ast} \psi^\alpha_{kl}(t) = \delta_{\alpha, \alpha'} \delta_{k, k'} \delta_{l, l'}.
  \label{s2:eq11}  
\end{equation}
By substituting Eqs.~\eqref{s2:eq8},~\eqref{s2:eq9} and~\eqref{s2:eq10} into~\eqref{s2:eq11}, it is straightforward to
show that $q_k$ can be given analytically as
\begin{eqnarray}
  q_k & = q_{[n,m]} = & \begin{cases}
    Q/e - n + 1, & \text{for $Q/e \in [n-1, n]$,}\\
    n + 1 - Q/e, & \text{for $Q/e \in (n, n+1]$,}\\
    0, & \text{otherwise.}
  \end{cases}\nonumber\\
  \label{s1:eq5}
\end{eqnarray}

The wave functions $\psi^\alpha_{kl}(t)$ can be regarded as Martin-Landauer-like wave packets
\cite{martin92_wave_packet_approac_to_noise}, which offers an intuitive way to interpret the time-resolved behavior of
the charged quasiparticles ($\alpha = c$) and $e$h pairs ($\alpha = e, h$). The decomposition of $G(t, t')$ can then be
given as
\begin{eqnarray}
  \label{s2:eq12}  
  && \hspace{0cm}i G(t, t') = \sum_{k,l} \psi^c_{kl}(t) [ \psi^c_{kl}(t') ]^{\ast} \\
  && \hspace{1.1cm}\mbox{}+ \sum_{k,l} \left[ \begin{tabular}{cc}
                                                    $\psi^e_{kl}(t)$, & $\psi^h_{kl}(t)$\\
                                                  \end{tabular}\right] \nonumber\\
  && \times \left[\begin{tabular}{cc}
                    $p_k$ & $i \sqrt{p_k[1 - p_k]}$\\
                    $-i \sqrt{p_k[1 - p_k]}$ & $1 - p_k$\\
                  \end{tabular}\right] \left[\begin{tabular}{c}
                                               $\psi^e_{kl}(t')$\\
                                               $\psi^h_{kl}(t')$\\
                                             \end{tabular}\right]^{\ast}.\nonumber
\end{eqnarray}

For wave packets carrying integer and fractional charges, both the charged quasiparticles and $e$h pairs can show
different natures, leading to wave functions with different features. To better demonstrate these differences, we shall
first concentrate on two concrete examples: wave packets carrying an unit ($Q=e$) and two-thirds ($Q/e=2/3$) electric
charges.

\section{Wave packet with unit charge}
\label{sec3}

Let us start our discussion from the wave packet carrying an unit electric charge ($Q=e$). In this case, the
decomposition of $G(t, t')$ takes a simple form:
\begin{equation}
  i G(t, t') = \sum_l \psi^c_{[1,0]l}(t) [ \psi^c_{[1,0]l}(t') ]^{\ast}.
  \label{s3:eq1}
\end{equation}
This indicates that the each wave packet contain only one charged quasiparticle associated with the index $k=[1,0]$. By
introducing the creation operator
\begin{equation}
  C^{\dagger}_{[1, 0]l} = \int^{+\infty}_{-\infty} dt \psi^c_{[1, 0]l}(t) \hat{a}^{\dagger}(t),
  \label{s3:eq2}
\end{equation}
the corresponding many-body state of the whole wave packet train can be expressed as
\begin{equation}
  | \mathbf{\Psi_{\rm train}} \rangle = \prod_{l=0, \pm 1, \pm 2, ...} C^{\dagger}_{[1, 0]l} | \mathbf{F}\rangle.
  \label{s3:eq3}
\end{equation}

\begin{figure}
  \includegraphics[width=7.5cm]{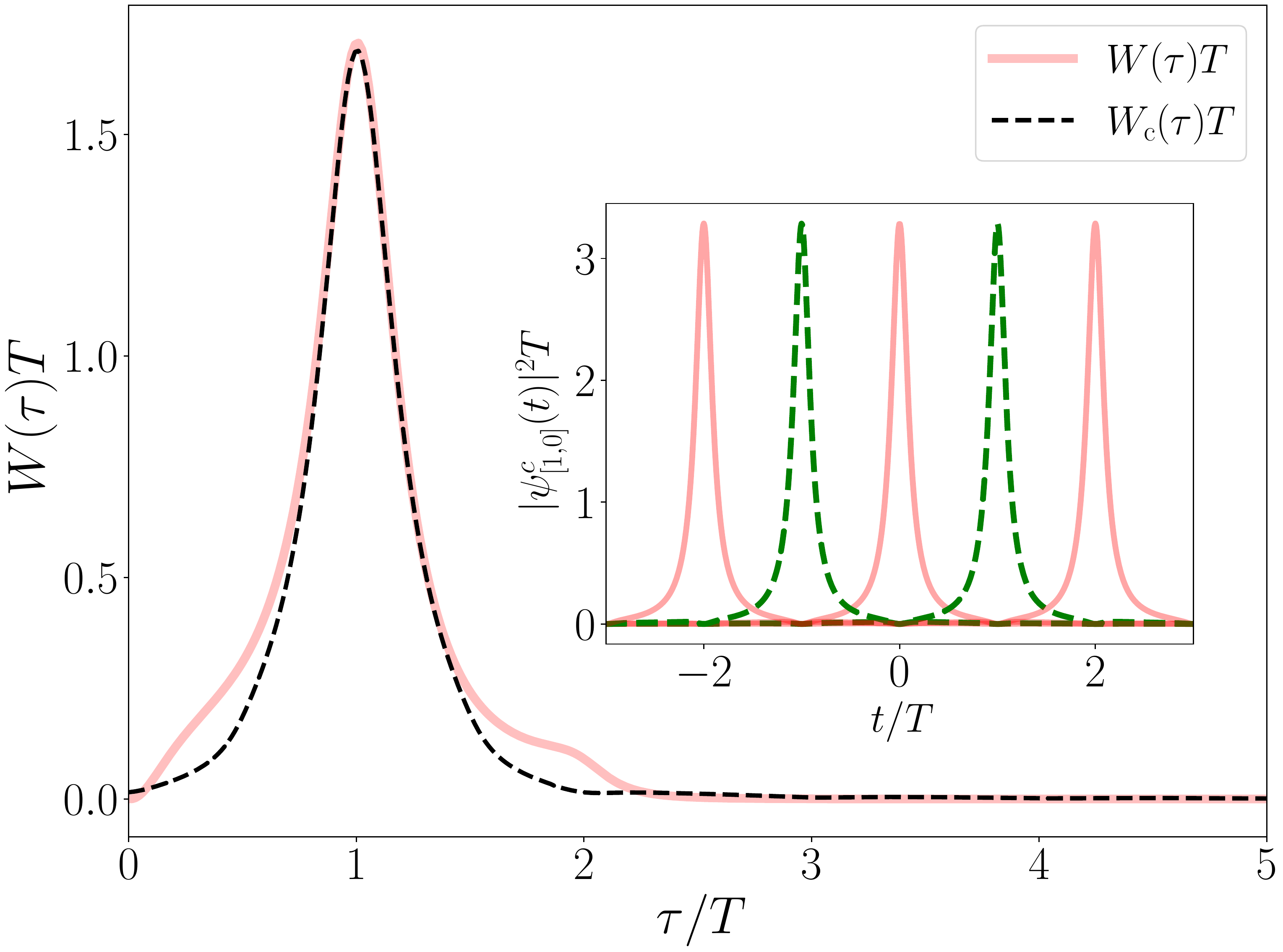}
  \caption{ (Color online) The WTD between electrons above the Fermi sea,
    corresponding to the width $W/T=0.1$. The red solid curve represents the
    exact WTD, while the black dashed curve represents the semi-classical
    approximation from Eq.~\eqref{s3:eq10}. The corresponding train of levitons
    are illustrated by the wave functions $|\psi^c_{[1, 0]l}(t)|^2T$ in the
    inset. The red solid curves correspond to $l=-2, 0$ and $2$, while the green
    dashed curves correspond to $l=-1$ and $1$.}
  \label{fig3}
\end{figure}

Equation~\eqref{s3:eq3} essentially corresponds to a periodic train of levitons. Accordingly, the wave functions
$\psi^c_{[1,0]l}(t)$ can be regarded as Martin-Landauer-like wave packets built from levitons. This can be seen more
clearly by carrying out the integration in Eq.~\eqref{s2:eq10} \footnote{Note that in this case, we have
  $q_{[1,0]}=1.0$.}:
\begin{equation}
  \psi^c_{[1, 0]l}(t) = U^c_{[1, 0]}(t) e^{-i \Omega (t-lT)/2} \sinc[\frac{\Omega (t-lT)}{2\pi}],
  \label{s3:eq5}
\end{equation}
where the periodic function
\begin{equation}
  U^c_{[1, 0]}(t) = \frac{\sqrt{ \cosh(\pi W/T) \sinh(\pi W/T)/T }}{\sin[\pi (t/T - i W/T)]},
  \label{s3:eq5-1} 
\end{equation}
represents the leviton train \cite{glattli-2016-hanbur-brown}. Each wave function $\psi^c_{[1,0]l}(t)$ exhibits a strong
peak around $t=lT$, corresponding to a leviton injected in the $l$-th period. Wave functions with different $l$ can form
a periodic sequence, providing an intuitive way to understand the structure of the wave packet train. This is
illustrated in the inset of Fig.~\ref{fig3}.

The wave functions $\psi^c_{[1, 0]l}(t)$ can provide an orthonormal basis set in the time domain, within which various
physical quantities can be expressed in a neat way. In particular, the current carried by the train of levitons can be
written as [see Appendix~\ref{app2} for details]
\begin{equation}
  I(t) = \sum_{l=0, \pm 1, \pm 2, ...} e| \psi^c_{[1, 0]l}(t) |^2.
  \label{s3:eq6} 
\end{equation}
One notices that in Eq.~\eqref{s3:eq6}, the current $I(t)$ is expressed as an incoherent summation of all the wave
functions $\psi^c_{[1, 0]l}(t)$, even if these functions can overlap with each other [see the inset of
Fig.~\ref{fig3}]. However, this does not mean that levitons contribute incoherently to the charge transport process. In
fact, the overlap between the wave functions can enhance the fluctuations of the waiting time between successive
electron injection. This effect can be seen more intuitively from the waiting time distribution (WTD) between electron
above the Fermi sea \cite{brandes08_waitin_times_noise_singl_partic_trans,
  albert11_distr_waitin_times_dynam_singl_elect_emitt, albert12_elect_waitin_times_mesos_conduc}.


The WTD can be calculated from the corresponding idle time probability $\Pi(t_s, t_e)$
\cite{dasenbrook16_quant_theor_elect_waitin_time_clock}. It can be expressed as the determinant [see Appendix~\ref{app3}
for details]:
\begin{equation}
  \Pi(t_s, t_e) = \det[ \hat{1} - \hat{Q}_{se} ],
  \label{s3:eq7} 
\end{equation}
where $\hat{1}$ denotes the unit operator and the operator $\hat{Q}_{se}$ counts the number of electrons injected in the
time interval $[t_s, t_e]$, whose energy is larger than the Fermi energy $E_F$. By introducing the Dirac notation
$\langle t | 1,0;l \rangle = \psi^c_{[1,0]l}(t)$, the matrix element of the operator $\hat{Q}_{se}$ can be given as
\begin{equation}
  \langle 1,0;l | \hat{Q}_{se} | 1,0;l' \rangle = \int^{t_e}_{t_s} dt
  [\psi^c_{[1,0]l}(t)]^{\ast} \psi^c_{[1,0]l'}(t).
  \label{s3:eq7-1} 
\end{equation}

For system under periodic driving, it is usually convenient to average the idle time probability over a single period:
\begin{equation}
  \Pi(\tau) = \int^{T/2}_{-T/2} dt_s \Pi(t_s, t_s + \tau).
  \label{s3:eq8} 
\end{equation}
In doing so, one obtains the time-averaged idle time probability $\Pi(\tau)$, which only depends on the length of the
time interval. The corresponding WTD can be given as
\begin{equation}
  W(\tau) = \langle \tau \rangle \partial^2_{\tau} \Pi(\tau),
  \label{s3:eq9} 
\end{equation}
with $\langle \tau \rangle$ being the mean waiting time.

The above equations offer a direct relation between the wave functions and WTD, where the overlap between levitons
manifest itself as the off-diagonal elements in Eq.~\eqref{s3:eq7-1}. When the overlap vanishes, the idle time
probability $\Pi(t_s, t_e)$ can be reduced to
\begin{equation}
  \Pi_c(t_s, t_e) = \prod_{l=0, \pm 1, \pm 2, ...} \Big[ 1 - \int^{t_e}_{t_s}dt |\psi^c_{[1,0]l}(t)|^2 \Big].
  \label{s3:eq10} 
\end{equation}
A quite similar result has been obtained for the ideal single-electron source built from the mesoscopic capacitor
\cite{hofer16_elect_waitin_times_mesos_capac}. The corresponding WTD $W_c(\tau)$ calculated from $\Pi_c(t_s, t_e)$ can
exhibit a strong peak around the point $\tau=T$ and drops rapidly to zero when $\tau > 2T$, as illustrated by the black
dashed curve in Fig.~\ref{fig3}. This indicates that one injects exactly one electron per period, corresponding to the
case of ideal single-electron injection. In realistic conditions, Eq.~\eqref{s3:eq10} can be regarded as a
semi-classical approximation. The presence of the overlap between levitons can lead to a deviation between the exact WTD
$W(\tau)$ and the semi-classical approximation $W_c(\tau)$. This can be seen by comparing the red solid curve
[$W(\tau)$] to the black dashed one [$W_c(\tau)$] in Fig.~\ref{fig3}, which are calculated for $W/T=0.1$. One can see
that the peak in the WTD is slightly broadened due to the overlap, indicating an enhancement of the fluctuations of the
waiting time.

\begin{figure}
  \includegraphics[width=7.5cm]{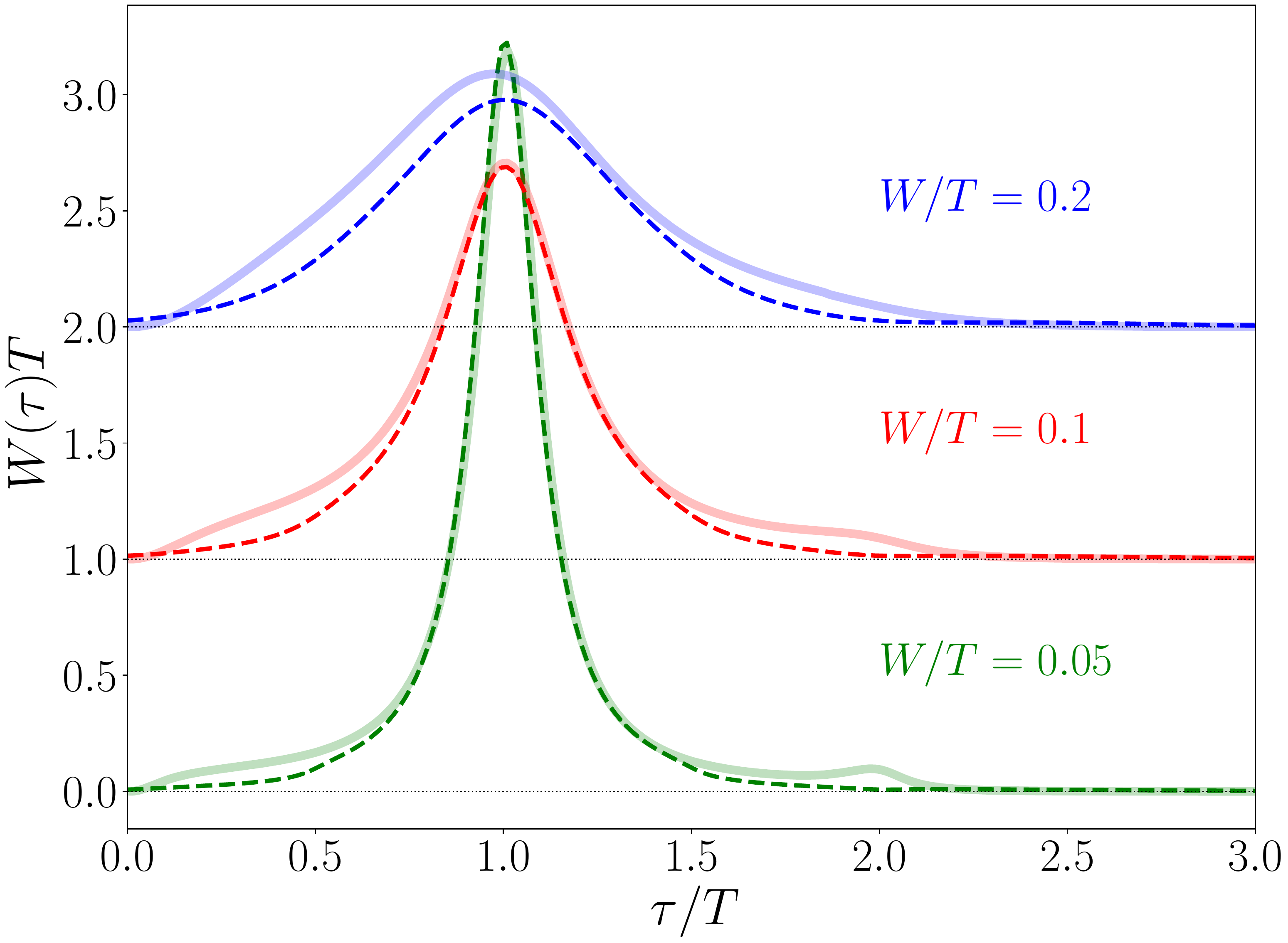}
  \caption{ (Color online) The WTDs between electrons above the Fermi sea, corresponding to $Q=e$ and the width
    $W/T=0.05$ (green), $0.1$ (red) and $0.2$ (blue). The solid curves represent the exact WTD, while the dashed curves
    represent the semi-classical approximation from Eq.~\eqref{s3:eq10}. Curves corresponding to different $W/T$ are
    shifted vertically for better visibility.}
  \label{fig4}
\end{figure}

In fact, the enhancement is not significant for $W/T=0.1$. Moreover, it can be suppressed by decreasing $W/T$. This is
illustrated in Fig.~\ref{fig4}, where we compare the WTDs for the width $W/T=0.05$, $0.1$ and $0.2$. This indicates that
the ideal single-electron injection can be approached in the limit $W/T \to 0$. Accordingly, the wave functions
$\psi^c_{[1,0]l}(t)$ are well-separated and can be treated as individual levitons in this limit.

The above results show that levitons can be well described by the single-body wave function $\psi^c_{kl}(t)$. In the
following section, we shall further demonstrate that the wave function $\psi^c_{kl}(t)$ can also be used to describe the
charged quasiparticles in the fractional-charged wave packet.

\section{Wave packet with two-thirds charges}
\label{sec4}

Now we turn to the wave packet carrying two-thirds electric charges
($Q/e=2/3$). In this case, each wave packet still contains only one charged
quasiparticle associated with the index $k=[1,0]$. Due to the dynamical
orthogonality catastrophe, one expect that the wave packet can also contain a
large amount of neutral $e$h pairs, when the pulse width $W/T$ is small
enough. However, as it is difficult to generate well-behaved voltage pulses with
too small width $W/T < 0.1$ \cite{dubois-2013-minim-excit,
  dubois-2013-integ-fract}, there exist only a rather limited number of $e$h
pairs under typical experimental conditions. In fact, even for the width
$W/T = 0.1$, we find that the excitation probabilities $p_k$ of the $e$h pairs
are all smaller than $0.15$.

\subsection{Charged quasiparticles}
\label{sec4a}

As a first step toward exploring the quantum state of the wave packets, let us omit the contribution of the $e$h pairs,
which is valid for large width $W/T$. In doing so, the correlation function $G(t,t')$ can be decomposed into the same
form as the one of levitons [see Eq.~\eqref{s3:eq1}]. However, the wave function $\psi^c_{[1,0]l}(t)$ takes a different
form, which can be written as
\begin{align}
  \psi^c_{[1, 0]l}(t) &= U^c_{[1, 0]}(t) \nonumber\\
  & \times \frac{e^{-i q_{[1,0]}\Omega (t-lT/q_{[1,0]})/2}}{\sqrt{q_{[1,0]}}} \sinc[\frac{q_{[1,0]}\Omega (t-lT/q_{[1,0]})}{2\pi}],
  \label{s4:eq1-1}
\end{align}
with the factor $q_{[1,0]} = 2/3$. By comparing the wave function of levitons in
Eq.~\eqref{s3:eq5}, we show that there are two differences between the two
cases: 1) The periodic function $U^c_{[1, 0]}(t)$ has to be obtained numerically
in this case; 2) While the function $U^c_{[1, 0]}(t)$ has the period $T$, the
sinc function in this case represents the wave packet localized around
$t=l(3T/2)$. This indicates that the wave functions $\psi^c_{[1,0]l}(t)$
correspond to the quasiparticles, which are injected with two different periods:
$T$ and $3T/2$. It is the double periodicity, which makes the quasiparticles
exhibit qualitatively different features from the ones of levitons.

\begin{figure}
  \includegraphics[width=7.5cm]{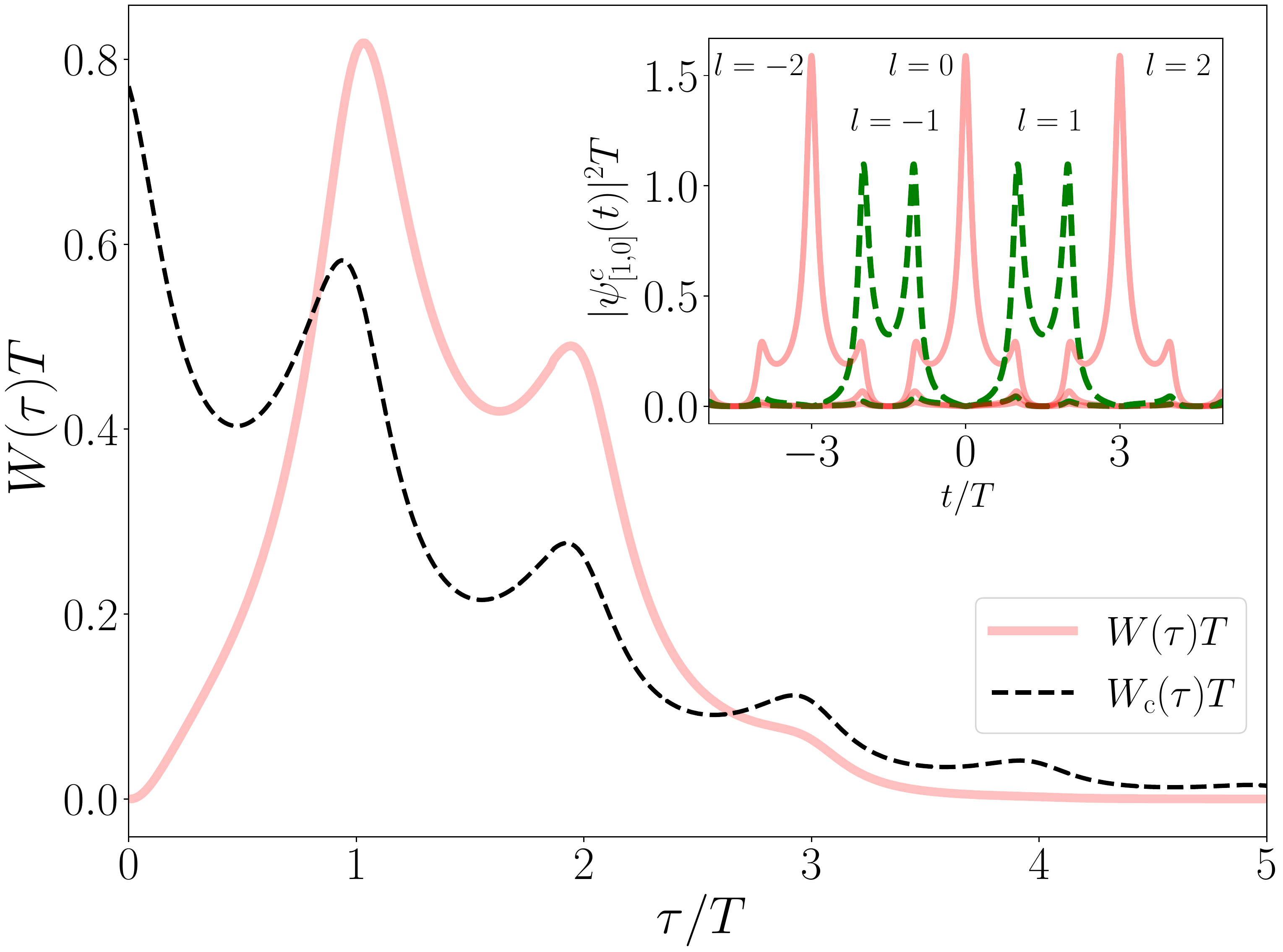}
  \caption{ (Color online) The WTD between electrons above the Fermi sea, corresponding to the width $W/T=0.1$. The
    corresponding train of charged quasiparticles are illustrated by the wave functions $|\psi^c_{[1, 0]l}(t)|^2T$ in
    the inset. The red solid curves correspond to $l=-2, 0$ and $2$, while the green dashed curves correspond to $l=-1$
    and $1$.}
  \label{fig5}
\end{figure}

The period $3T/2$ decides the charges carried by the quasiparticles. In fact, as
the wave functions $\psi^c_{[1, 0]l}(t)$ with different $l$ are still orthogonal
to each other [see Eq.~\eqref{s2:eq11}], one can still express the current as
the incoherent summation of them, which has the same form as the one of levitons
[see Eq.~\eqref{s3:eq6}]. However, as these wave functions are separated from
each other by the time interval $3T/2$ [see the inset of Fig.~\ref{fig5}], on
average each quasiparticle can carry only $2e/3$ charge within a single period
$T$, making them behave {\em effectively} like quasiparticles carrying
fractional charges.

Note that in this case, the wave functions can exhibit two different
profiles. For $l=-2, 0$ and $2$ (red solid curves), the wave functions
$\psi^c_{[1, 0]l}(t)$ can exhibit a strong peak, which is accompanied by two
small shoulder peaks. In contrast, for $l=-1$ and $1$ (green dashed curves), the
wave functions $\psi^c_{[1, 0]l}(t)$ exhibit double peak structures. This is a
direct consequence of the double periodicity of the wave functions. In fact,
from Eq.~\eqref{s4:eq1-1}, one can see that when the periods corresponding to
$U^c_{[1, 0]}(t)$ (with the period $T$) and the sinc function (with the period
$T/q_{[1,0]}$) do not match, for $q_{[1,0]} = A/B$ (with $A$ and $B$ being
coprime integers), the wave functions can exhibit $A$ different profiles, which
are separated from each other with the extended period $BT/A$.

Due to the mismatch between the two periods, the wave functions are strongly
overlapped with each other. This can be seen intuitively from the inset of
Fig.~\ref{fig5}. The overlap can induce a large fluctuation of the waiting time,
which can be seen from the corresponding WTD \footnote{Generally speaking, the
  electron component of the $e$h pairs can also contribute to the WTD. However,
  the contribution remains negligible due to the small excitation probability
  for the width $W/T > 0.05$.}. This is illustrated by the red solid curves in
the main panel of Fig.~\ref{fig5}. One can see that the WTD exhibits a series of
peaks at multiplies of the repetition period $T$. This indicates the presence of
the cycle-missing event, in which the voltage pulse fails to inject an electron
within a single period $T$ \cite{potanina17_elect_waitin_times_period_driven,
  hofer16_elect_waitin_times_mesos_capac}.

\begin{figure}
  \includegraphics[width=7.5cm]{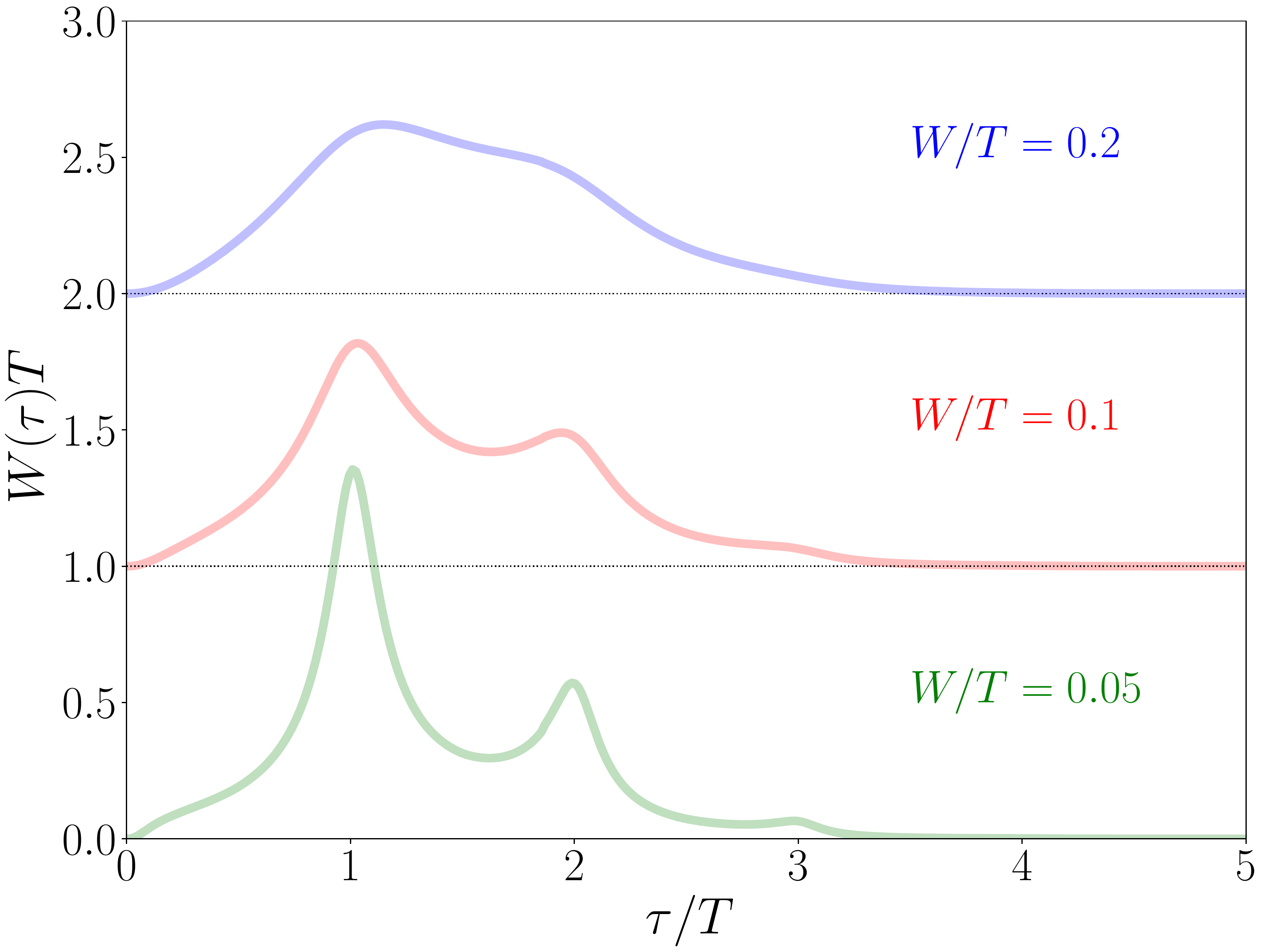}
  \caption{ (Color online) The WTD between electrons above the Fermi sea, corresponding to $Q/e=2/3$ and the width
    $W/T=0.05$ (green), $0.1$ (red) and $0.2$ (blue). Curves corresponding to different $W/T$ are shifted vertically for
    better visibility.}
  \label{fig6}
\end{figure}

As the overlap between the wave functions are rather large, the semi-classical approximation $W_c(\tau)$ of the WTD
[Eq.~\eqref{s3:eq10}] is inapplicable. One can see that $W_c(\tau)$ largely overestimates the WTD around the point
$\tau=0$, which is illustrated by the black dashed curve in Fig.~\ref{fig5}. In fact, $W_c(\tau)$ gives an unphysical
value around this point: The WTD should be zero at $\tau=0$ due to the Pauli principle. Unlike the case of levitons, the
overlap between the wave functions cannot be eliminated by just decreasing the width $W/T$. As a consequence, the
multiple-peak structure of the WTD preserves as $W/T$ decreases. This is illustrated in Fig.~\ref{fig6}, corresponding
to $W/T=0.2$, $0.1$ and $0.05$.

The above results explains the nature of the fractional-charged quasiparticles:
they are just quasiparticles injected with an extended period $T/q_k$, which is
longer than the period $T$ of the driving pulses. The wave functions of these
quasiparticles are always strongly overlapped with each other, manifesting
themselves at multiple peaks in the corresponding WTD. The feature of these
quasiparticles can be characterized by the factor $q_k$, indicating that each
quasiparticle can carry $e q_k$ charges per period $T$, making them behave {\em
  effectively} as fractional-charge quasiparticles.


\subsection{Electron-hole pairs}
\label{sec4c}

\begin{figure}
  \includegraphics[width=7.5cm]{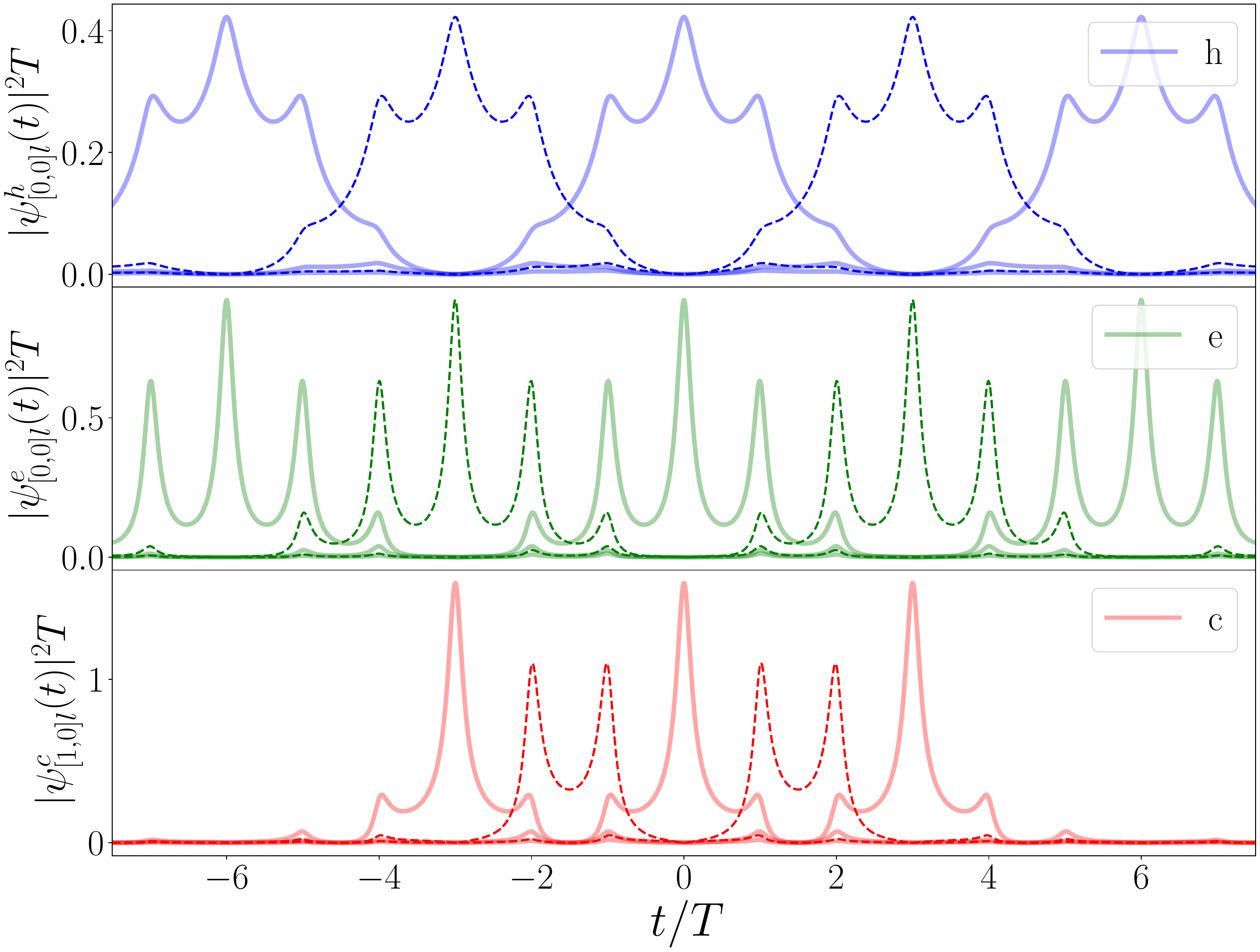}
  \caption{ (Color online) Wave functions of the charged quasiparticle
    ($k = [1, 0]$) and the $e$h pair ($k = [0, 0]$). The red curves represent
    the wave functions of the charged quasiparticle. The green and blue curves
    represent the wave functions for the electron and hole components of the
    $e$h pair, respectively. All the wave functions are calculated with
    $W/T=0.1$. The solid (dashed) curves from left to right correspond to
    $l=-2$, $0$ and $2$ ($l=-1$ and $1$), respectively.}
  \label{fig7}
\end{figure}

Now let us briefly discuss the $e$h pairs in the wave packet. For $W/T=0.1$, we
find that each wave packet contains only one $e$h pair, which is associated with
the index $k=[0, 0]$. The corresponding excitation probability $p_{[0,0]}$ is
only $0.138$. The other $e$h pairs are negligible due to their small excitation
probabilities \footnote{They are all smaller than $0.002$}. The $e$h pairs can
be described in a similar way as the charged quasiparticles. In fact, the wave
functions of the electron and hole components can be expressed in a similar form
as shown in Eq.~\eqref{s4:eq1-1}:
\begin{align}
  \psi^{e/h}_{[0, 0]l}(t) &= U^{e/h}_{[0, 0]}(t) \nonumber\\
  & \times \frac{e^{-i q_{[0,0]}\Omega (t-lT/q_{[0,0]})/2}}{\sqrt{q_{[0,0]}}} \sinc[\frac{q_{[0,0]}\Omega (t-lT/q_{[0,0]})}{2\pi}].
    \label{s4:eq9}
\end{align}
with the factor $q_{[0,0]}=1/3$. The corresponding wave functions
$\psi^{e/h}_{[1,0]l}(t)$ are plotted by the green/blue curves in
Fig.~\ref{fig7}, where the wave function $\psi^c_{[1,0]l}(t)$ of the charged
quasiparticles are also plotted by the red curves for comparison. One can see
that in this case, the wave functions for the electron (hole) component exhibit
only one type of profiles. They are separated from each other by the time
interval $3T$, making them behave as quasiparticles carrying $e/3$ charges. Note
that electron and hole components carry the same amount of charges but with
opposite sign, which cannot contribute to the total charge $Q$ of the wave
packet.

By combining the information of both the charged quasiparticles and $e$h pairs,
the quantum state of the whole train of wave packet can be written as
\begin{align}
  \label{s4:eq10}  
  | \mathbf{\Psi_{\rm train}} \rangle &= \prod_{l=0, \pm 1, \pm 2, ...} C^{\dagger}_{[1, 0]l} \Big[\sqrt{1 - p_{[0, 0]}} \nonumber\\
                                      &\hspace{1cm}+ i \sqrt{p_{[0, 0]}}(B^e_{[0, 0]l})^{\dagger} (B^h_{[0, 0]l})^{\dagger} \Big] | \mathbf{F}\rangle,
\end{align}
with
\begin{align}
  C^{\dagger}_{[1,0]l} & = \int^{+\infty}_{-\infty} dt \psi^c_{[1,0]l}(t) \hat{a}^{\dagger}(t), \nonumber\\
  (B^e_{[0,0]l})^{\dagger} & = \int^{+\infty}_{-\infty} dt \psi^e_{[0,0]l}(t) \hat{a}^{\dagger}(t), \nonumber\\
  (B^h_{[0,0]l})^{\dagger} & = \int^{+\infty}_{-\infty} dt \psi^h_{[0,0]l}(t) \hat{a}(t).
  \label{s4:eq10-1}  
\end{align}
This provides a full information of the injected electric wave packet. It allows
us to elucidate how the quantum state of wave packets can evolve as the flux of
the pulses changes. In the following section, we shall concentrate on the
evolution of the charged quasiparticles. We shall show how levitons can emerge
as the flux approaches an integer value.

\section{Evolution of charged quasiparticle}
\label{sec5}

\begin{figure}
  \includegraphics[width=8.5cm]{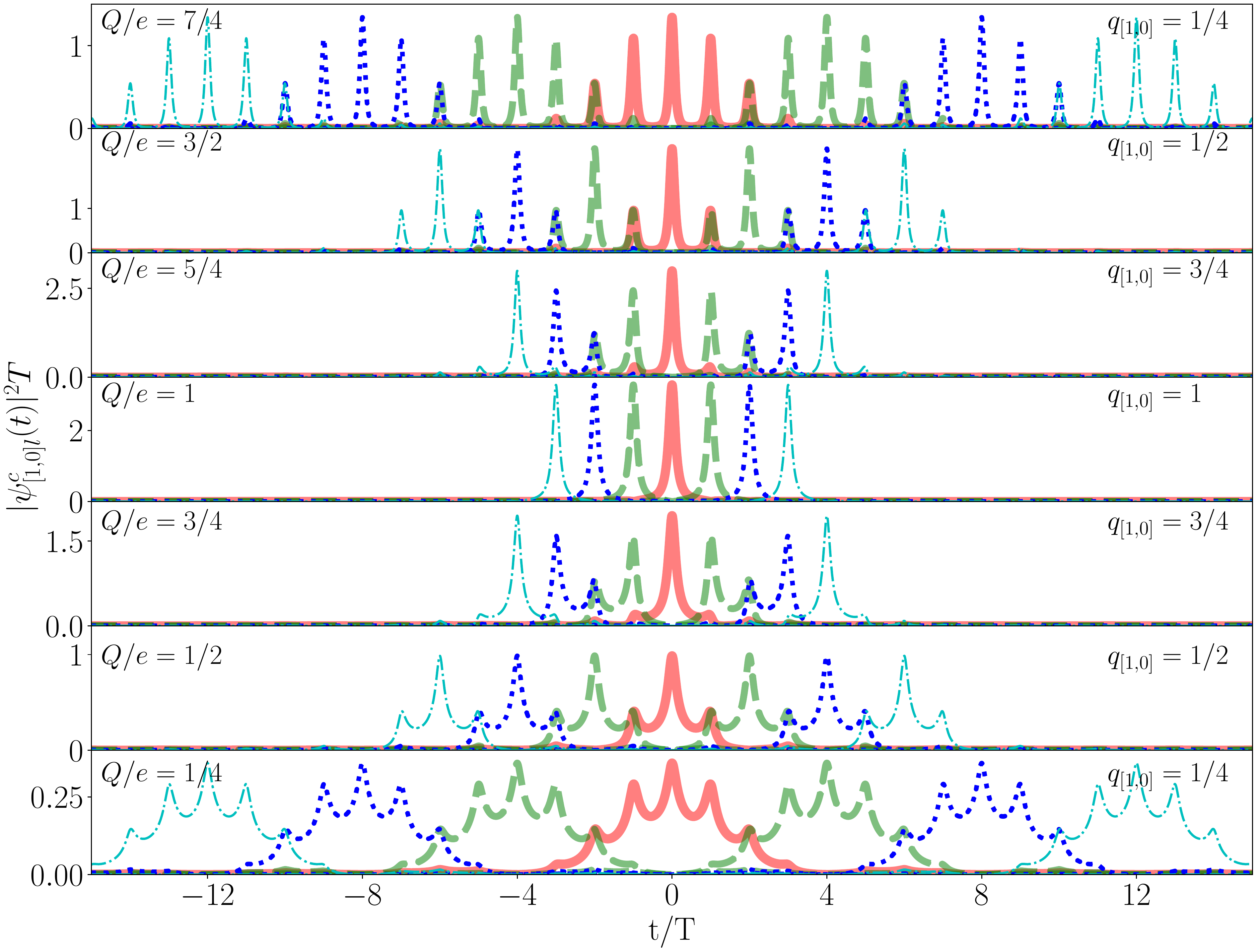}
  \caption{ (Color online) Wave functions of the charged quasiparticle
    $|\psi^c_{[1,0]l}(t)|^2T$, corresponding to $W/T=0.1$ and
    $Q/e \in (0.0, 2.0)$. For each value of $Q/e$, we plot the wave functions
    for $l=-3 \sim 3$. Curves with different colors and line types correspond to
    $|\psi^c_{[1,0]l}(t)|^2T$ with different $l$.}
  \label{fig8}
\end{figure}

The evolution of the charged quasiparticles can be fully described by the
single-body wave function $\psi^c_{kl}(t)$. This is illustrated in
Fig.~\ref{fig8}, corresponding to the index $k=[1,0]$. In the figure, we choose
$W/T=0.1$ and $Q/e \in (0.0, 2.0)$. Curves with different colors and line types
correspond to wave functions $\psi^c_{[1,0]l}(t)$ with different $l$. As the
factor $q_{[1,0]}$ can play an important role, we also show the corresponding
$q_{[1,0]}$ alongside the wave functions.

From the figure, one first notices that one has $q_{[1,0]} = Q/e$ when
$Q/e \in (0.0, 1.0)$. For $q_{[1,0]} = 1/4$, all the wave functions of the
quasiparticles exhibit the same profile. These quasiparticles are injected with
the extended period $4T$, indicating that they can carry $e/4$ charge within
each period $T$. As $q_{[1,0]}$ increases from $1/4$ to $1/2$, the extended
period is reduced to $2T$, indicating that the quasiparticles evolve into the
$e/2$-charged quasiparticles. As $q_{[1,0]}$ further increases from $1/2$ to
$3/4$, there can exist three types of quasiparticle, which are injected with the
extended period $3T/4$, leading to $3e/4$ charges per period. As $q_{[1,0]}$
reaches $1.0$, all the quasiparticles can evolve into levitons, which are
injected with the period $T$.

\begin{figure}
  \includegraphics[width=7.0cm]{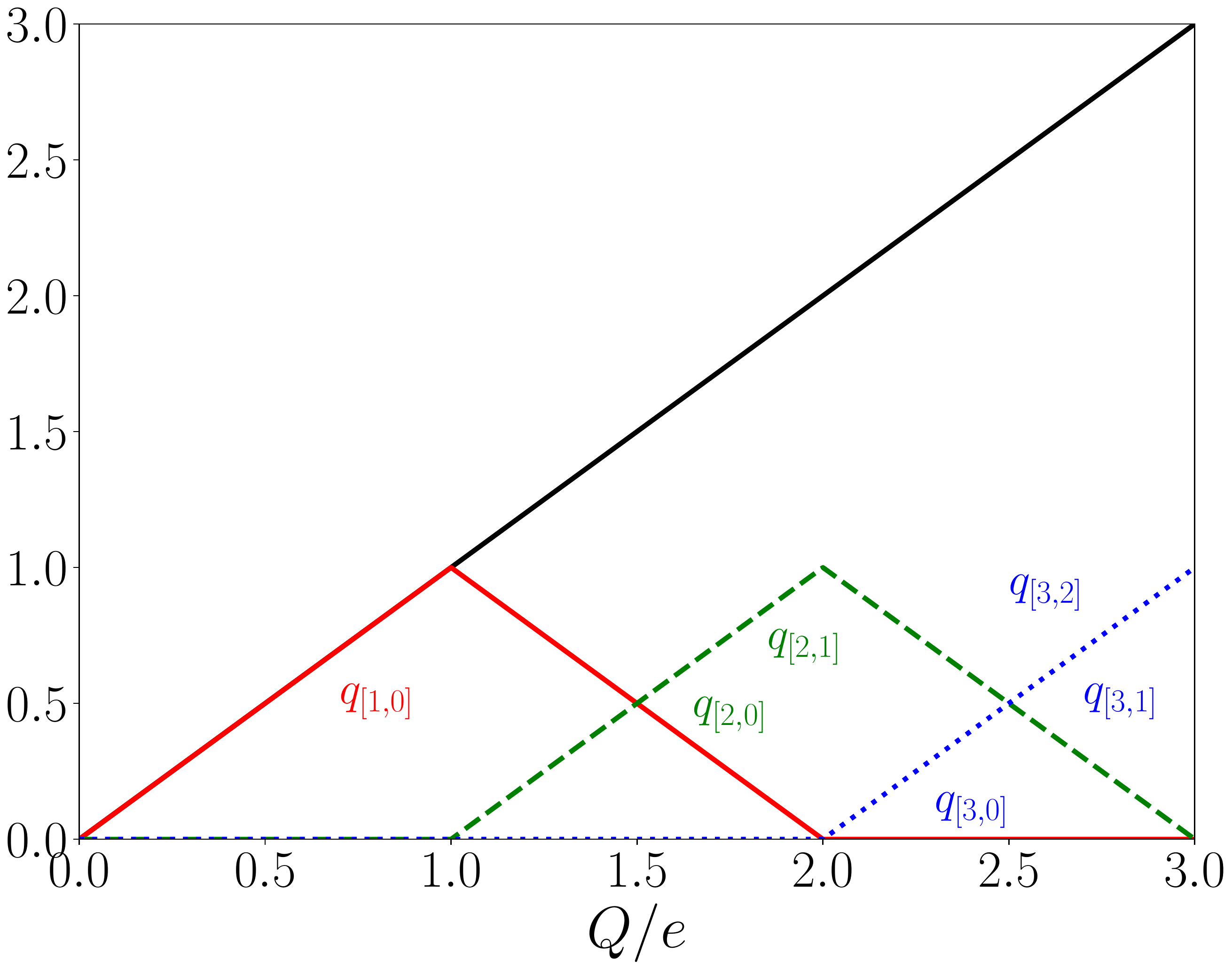}
  \caption{ (Color online) Factor $q_k$ as functions of $Q/e$. The red solid curve represents $q_{[1,0]}$. The green
    dashed curves represent $q_{[2,0]}$ and $q_{[2,1]}$. Note that one has $q_{[2,0]} = q_{[2,1]}$, so the two curves
    are overlapped. Similarly, the blue dotted curves represent $q_{[3,0]}$, $q_{[3,1]}$ and $q_{[3,2]}$. The black
    solid curve represents the charge of the wave packet $Q/e$, which satisfies $Q = e\sum_k q_k$.}
  \label{fig9}
\end{figure}

For $Q/e \in (1.0, 2.0)$, one has $q_{[1,0]} = 2 - Q/e$. As $Q/e$ increases in
this region, $q_{[1,0]}$ is dropping linearly to zero. Accordingly, the levitons
can evolve back into fractional-charged quasiparticles, which are injected with
the extended period $T/q_{[1,0]}$. Note that one has $T/q_{[1,0]} \to +\infty$
for $q_{[1,0]} \to 0$. This implies that the corresponding quasiparticles cannot
be injected in this limit, since the time interval between successive
quasiparticle injection tends to infinity. Generally speaking, quasiparticles
associated with the index $k=[n,m]$ can only be injected when
$Q/e \in [n-1, n+1]$, as shown in Eq.~\eqref{s1:eq5}. This can be seen more
clearly for Fig.~\ref{fig9}.

\begin{figure}
  \includegraphics[width=7.5cm]{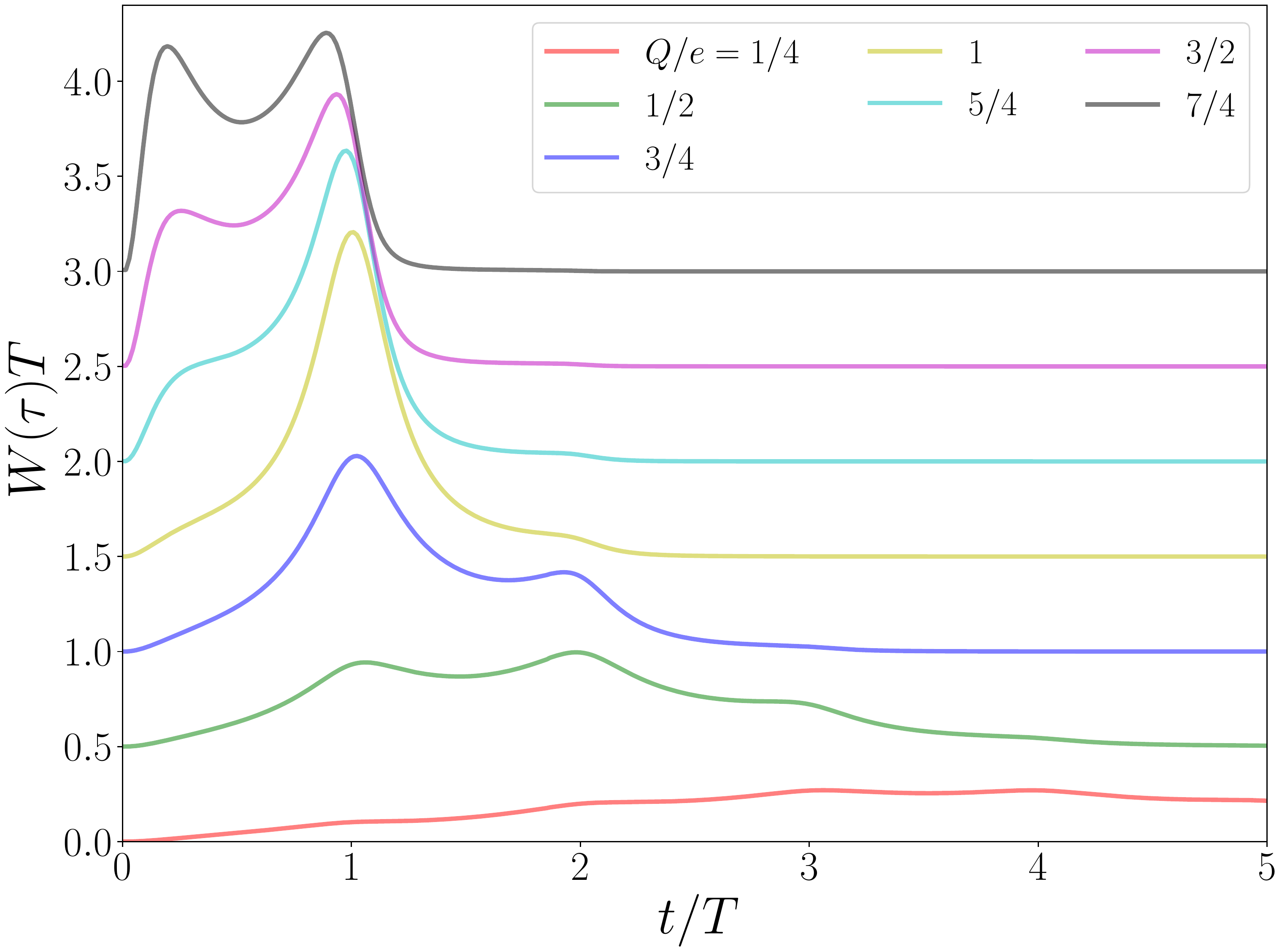}
  \caption{ (Color online) The WTDs between electrons above the Fermi sea, corresponding to $Q/e \in (0.0, 2.0)$ and the
    width $W/T=0.1$. Curves corresponding to different $Q/e$ are shifted vertically for better visibility. }
  \label{fig10}
\end{figure}

\begin{figure}
  \includegraphics[width=8.5cm]{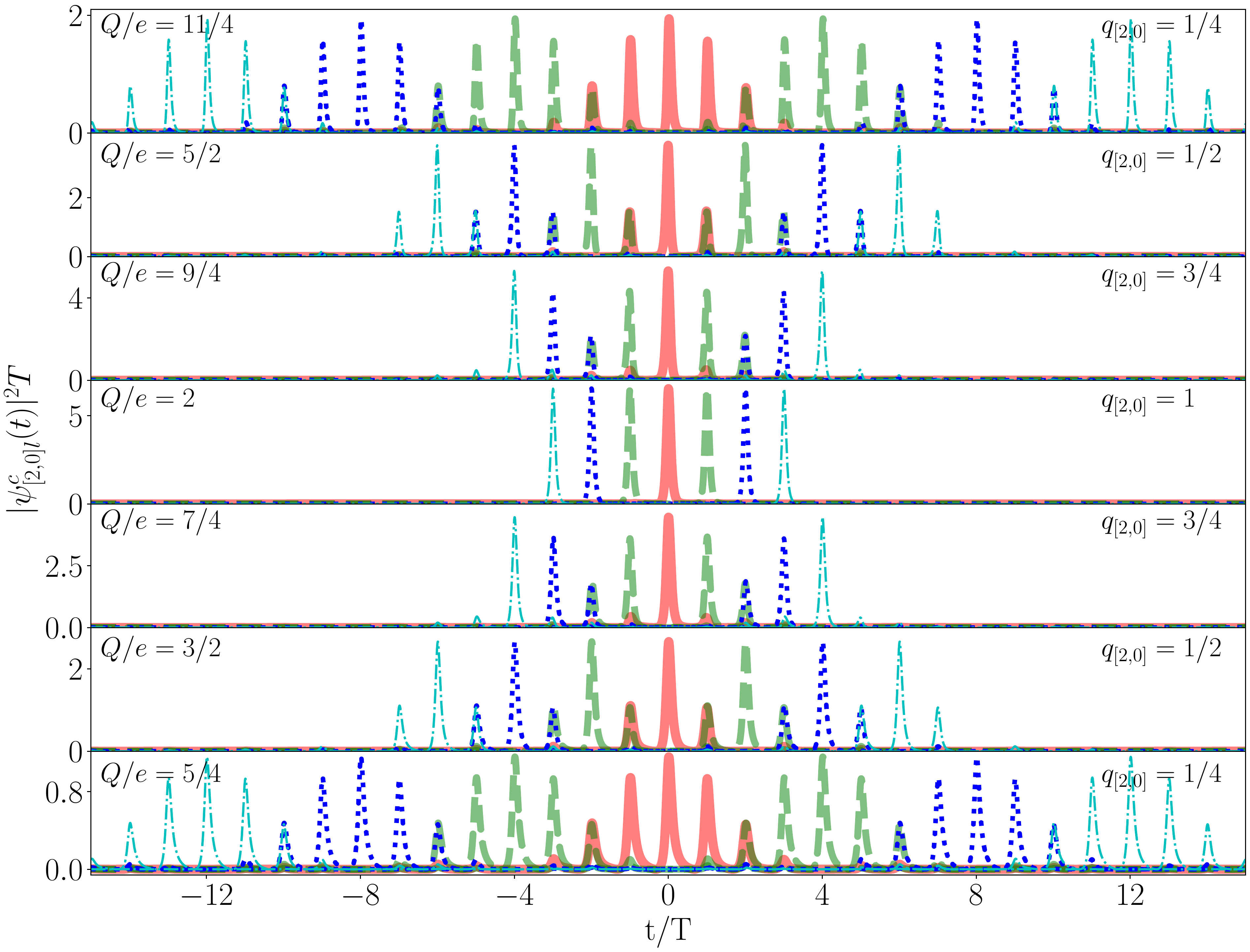}
  \includegraphics[width=8.5cm]{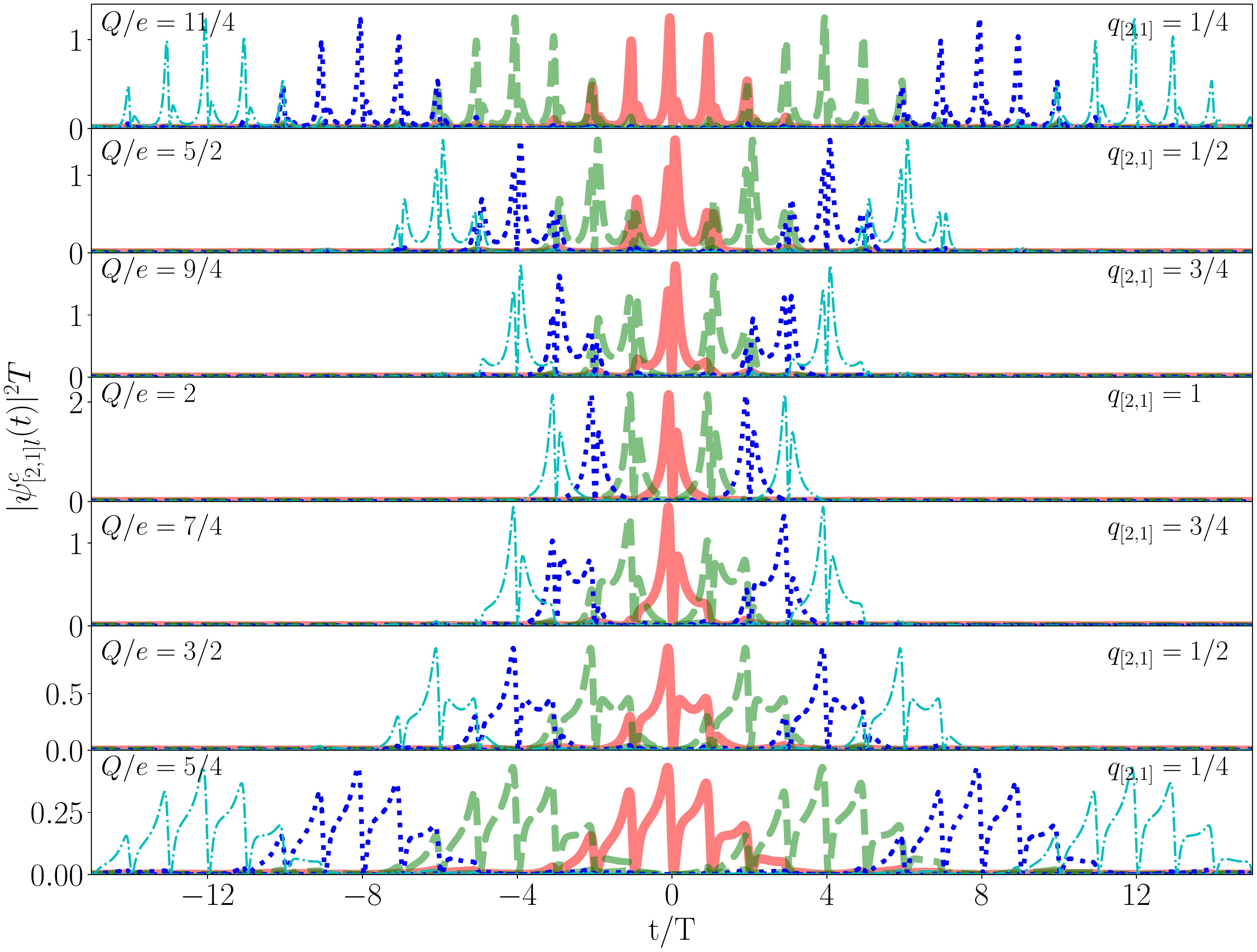}
  \caption{ (Color online) (Color online) Wave functions of the charged
    quasiparticle $|\psi^c_{[2,0]l}(t)|^2T$ (a) and $|\psi^c_{[2,1]l}(t)|^2T$
    (b), corresponding to $W/T=0.1$ and $Q/e \in (1.0, 3.0)$. For each value of
    $Q/e$, we plot the wave functions for $l=-3 \sim 3$. Curves with different
    colors and line types correspond to the wave functions with different $l$. }
  \label{fig11}
\end{figure}

The evolution of the quasiparticles can also be seen from the corresponding WTD,
as illustrated in Fig.~\ref{fig10}. One can see that for $Q/e = 1/4$, the
waiting time has a rather wide distribution. This is because the corresponding
wave functions of the quasiparticles are strongly overlapped, as shown in
Fig.~\ref{fig8}. As $Q/e$ approaches $1.0$, the WTD $W(\tau)$ tends to exhibit a
strong peak around $\tau = T$, indicating the emergence of levitons. Hence the
evolution of the wave functions for $Q/e < 1.0$ can also be tracked by using the
corresponding WTD. As $Q/e$ goes above $1.0$, additional charged quasiparticles
can be injected. From Fig.~\ref{fig9}, one can see that two additional
quasiparticles $k=[2,0]$ and $[2,1]$ can emerge. The evolution of these two
quasiparticles are demonstrated in Fig.~\ref{fig11}. By comparing to
Fig.~\ref{fig8}, one can see that they evolves in a similar way as the
quasiparticle $k=[1,0]$. As these two quasiparticles can also contribute to the
WTD, it is difficult to read the evolution of a single charged quasiparticles
from the WTD when $Q/e > 1.0$.

\section{Evolution of electron-hole pairs and shot noise}
\label{sec6}

As levitons evolve into fractional-charged quasiparticle, additional $e$h pairs
can be excited. Due to the small excitation probabilities, the $e$h pairs can
have little contribution to the WTD between electrons above the Fermi sea
\footnote{Generally speaking, the electron component of the $e$h pair can also
  contribute to the WTD between electrons above the Fermi sea. However, the
  contribution remains negligible for $W/T =0.05$.}. In contrast, it can have
pronounced impact on the shot noise, which has been extensively studied in
previous works \cite{dubois-2013-integ-fract, bocquillon-2014-elect-quant,
  vanevic-2012-contr-elect, vanevic-2016-elect-elect}. When the wave packet is
partitioned at a localized scatter with transmission probability $D$, both the
charged quasiparticles and $e$h pairs can contribute to the shot noise $S_N$. It
can be decomposed into two part [see Appendix~\ref{app2} for details]:
$S_N = S_c + S_{ex}$, where
\begin{align}
  S_c &= S_0 \sum_k q_k, \nonumber\\
  S_{ex} &= 2 S_0 \sum_k q_k p_k .
           \label{s6:eq1}
\end{align}
with $S_0 = 2 \frac{e^2}{h} D(1-D)\hbar\Omega$ being the typical scale of the shot noise.

\begin{figure}
  \includegraphics[width=7.0cm]{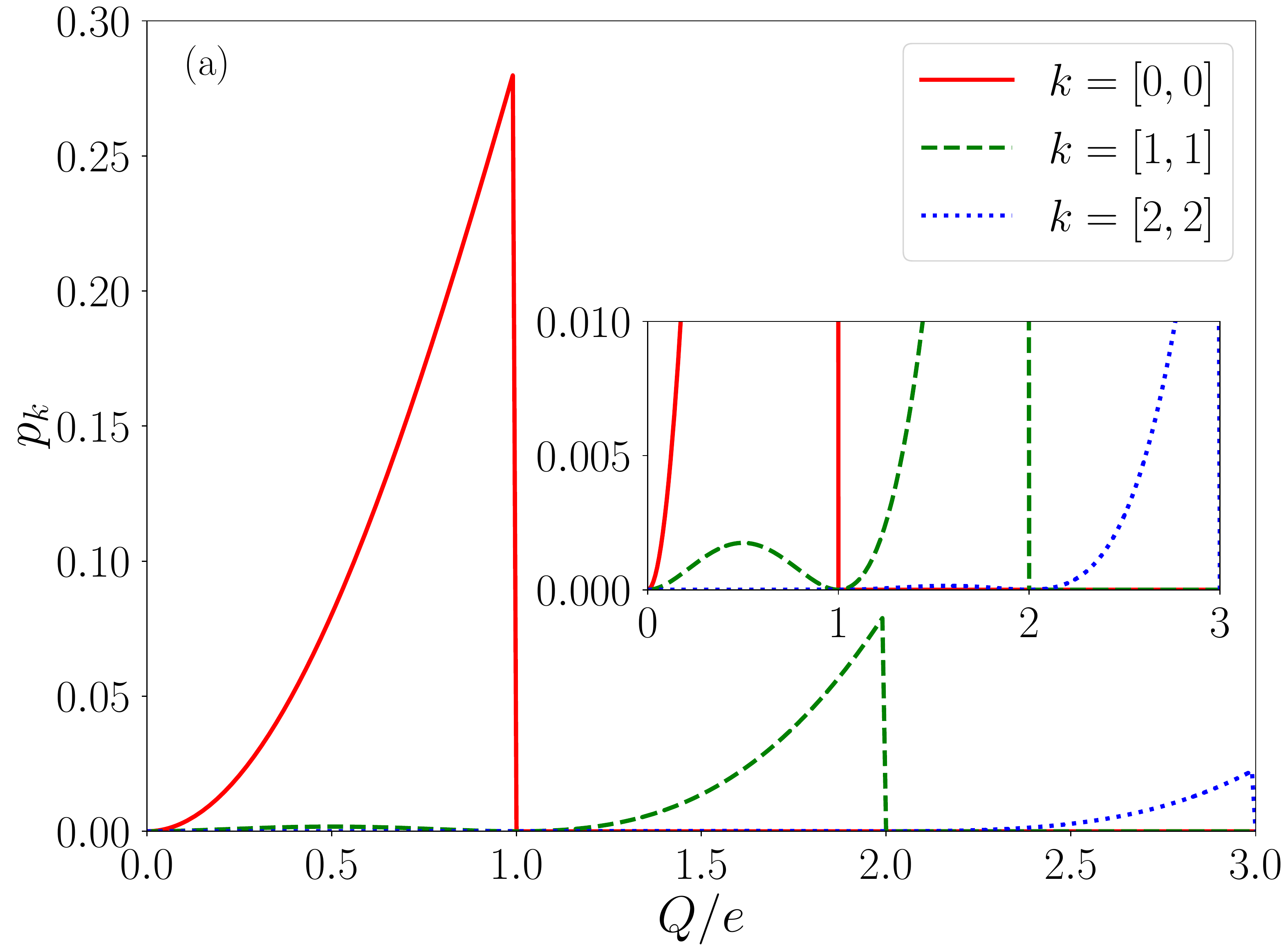}
  \includegraphics[width=7.0cm]{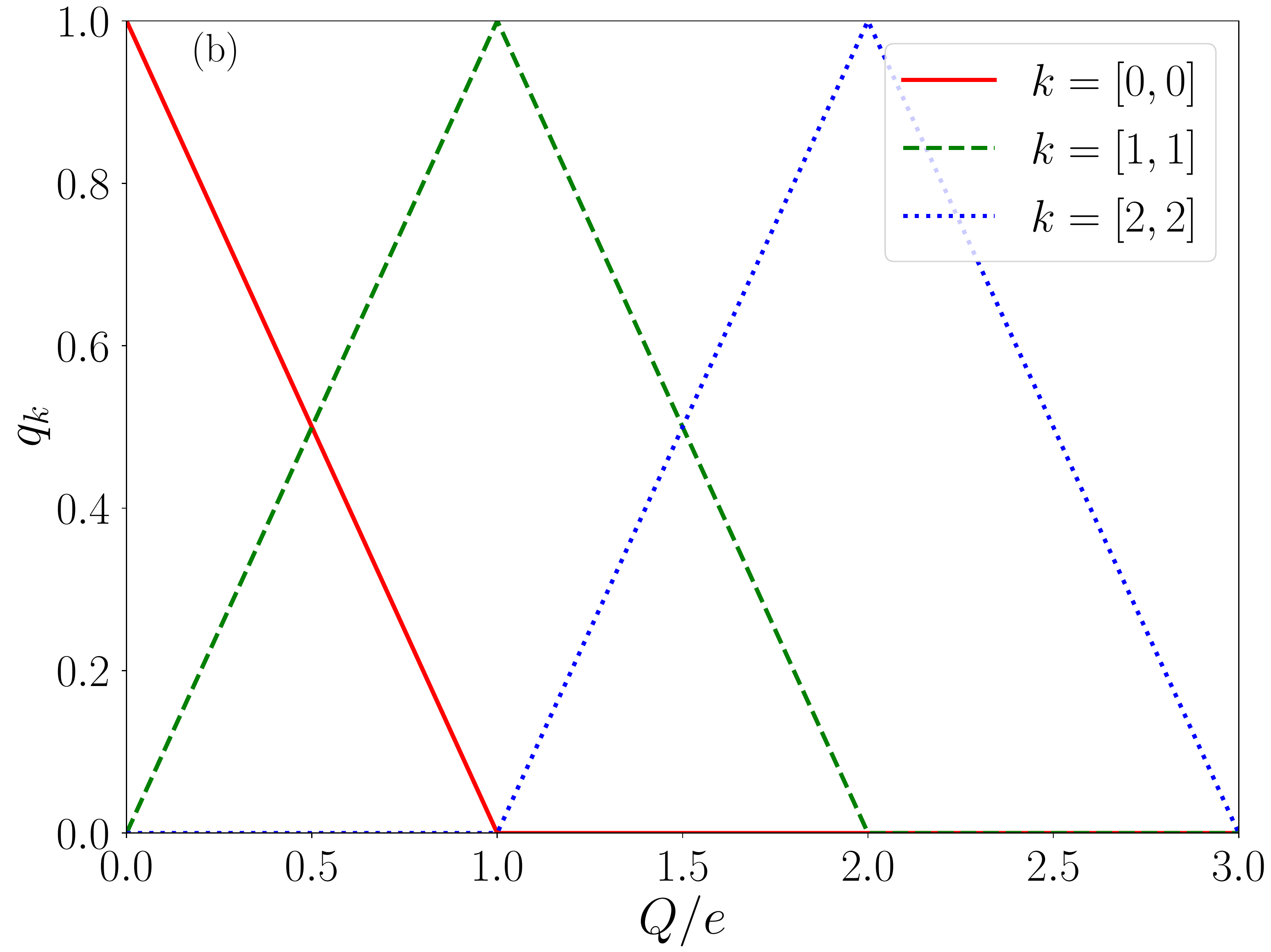}
  \includegraphics[width=7.0cm]{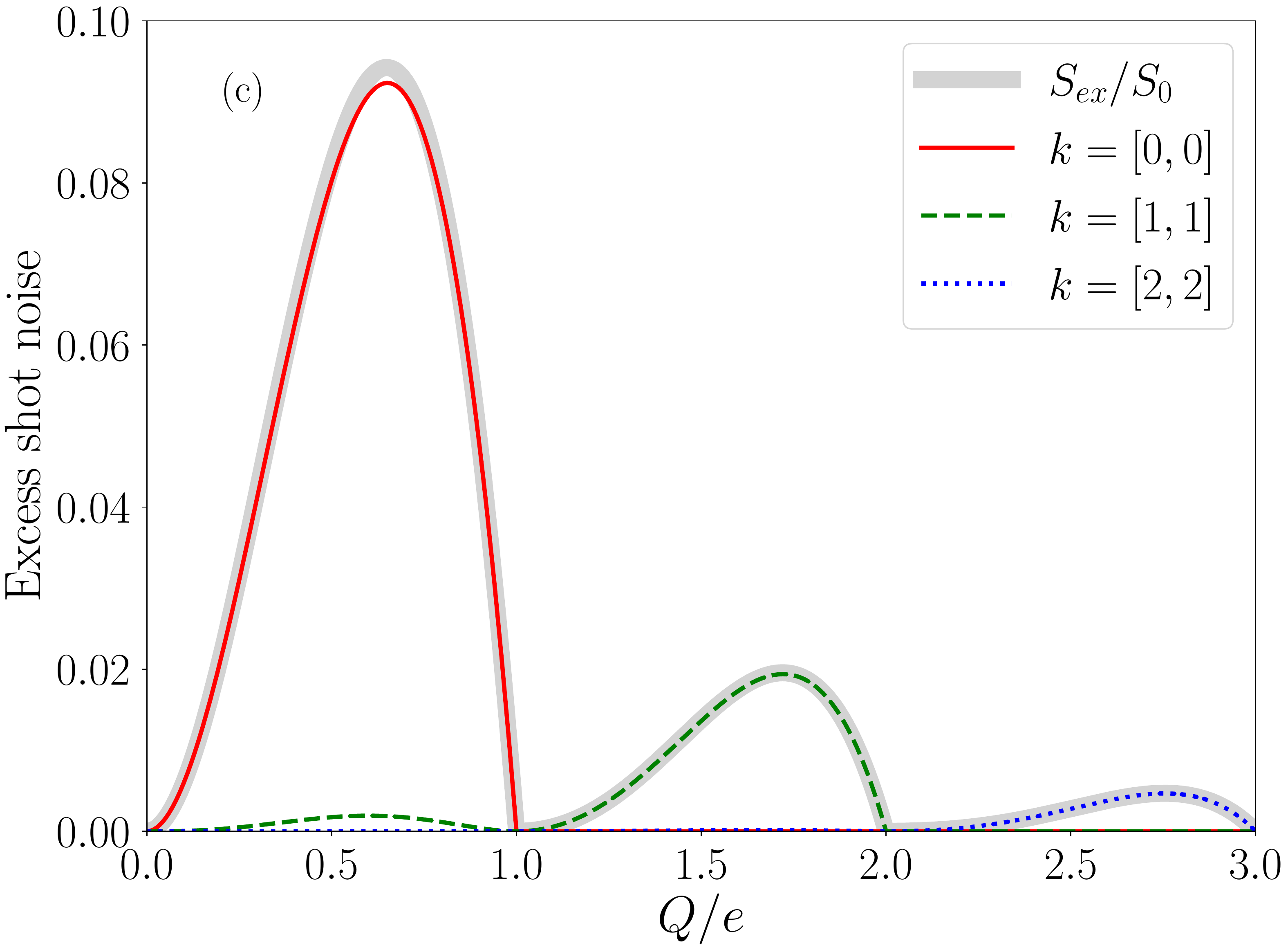}
  \caption{ (Color online) (a) The excitation probabilities $p_k$ for the $e$h
    pairs. The inset shows the zoom-in of the figure. The $p_k$ of other $e$h
    pairs are too small to be seen from the figure. (b) The injection
    probabilities $q_k$ of the electron/hole component of the $e$h pairs. (c)
    Excess shot noise $S_{ex}$ as a function of the charge $Q/e$ of the wave
    packet. The red solid, green dashed and blue dotted curves represent the
    contribution from the $e$h pairs $k=[0, 0]$, $k=[1, 1]$ and $k=[2, 2]$,
    respectively. The excess shot noise is normalized to
    $S_0 = 2 \frac{e^2}{h} D(1-D)\hbar\Omega$.}
  \label{fig12}
\end{figure}

The first part corresponds to the contribution of the charged quasiparticles. It
is solely decided by the charge $Q$ of the wave packet, since one has
$\sum_k q_k = Q/e$ from Eq.~\eqref{s1:eq5}. The second part is the excess shot
noise, which has been used extensively to characterize the feature of $e$h pairs
\cite{dubois-2013-integ-fract, vanevic-2007-elemen-event}. By using the
information of the excitation probability $p_k$ and the factor $q_k$, one can
decompose the excess shot noise $S_{ex}$ into the contribution of individual
$e$h pairs. This is illustrated in Fig.~\ref{fig12}. From the figure, one can
identify the contribution of three $e$h pairs, corresponding to $k=[0,0]$,
$k=[1,1]$ and $k=[2,2]$. These $e$h pairs dominate the excess shot noise
$S_{ex}$ in different regions. Such decomposition makes it possible to extract
the information of individual $e$h pairs from the excess shot noise. By
combining the WTD with the shot noise, one can hence obtain the full information
of the evolution of the quantum state of the wave packet.

\section{Summary and Outlook}
\label{sec7}

In summary, we have present a general approach to extract the quantum state of
wave packets injected by Lorentzian pulse train with arbitrary flux. We show
that the charged quasiparticles can be described by a set of single-body wave
functions $\psi^c_{kl}(t)$. These wave functions can be regarded as
Martin-Landauer-like wave packets, which offers an intuitive way to interpret
their time-resolved behaviors. In integer-charged wave packets, the charged
quasiparticles are levitons, which are injected with the same period as the
pulse train. No $e$h pairs can be injected in this case. In fractional-charged
wave packets, the charged quasiparticles can be injected with two different
periods. Due to the double periodicity, their wave functions can exhibit
different profiles. They can form a periodic train, whose period is longer than
the period of the pulse train. This makes them behave effectively as
quasiparticles carrying fractional charges. We show that the evolution of the
charged quasiparticles can be seen from the WTD between electrons above the
Fermi sea. Our approach can also be used to describe the evolution of $e$h
pairs, which can be tracked by using the shot noise. Note that although our
approach is demonstrated for the Lorentzian pulses, it is rather general and can
be applied to pulses with arbitrary profiles. We expect our work will be helpful
to explore the full potential of the voltage electron source.

\begin{acknowledgments}
  The authors would like to thank Professor D. C. Glattli and Professor M. V.
  Moskalets for helpful comments and discussion. This work was supported by the
  National Key R\&D Program of China under Grant No. 2016YFF0200403, the Key
  Program of National Natural Science Foundation of China under Grant
  No. 11234009, Young Scientists Fund of National Natural Science Foundation of
  China under Grant No. 11504248 and SCU Innovation Fund under Grant
  No. 2020SCUNL209.
\end{acknowledgments}

\appendix

\section{Bloch-Messiah reduction within the framework of scattering matrix
  theory}
\label{app1}

The basic idea is straightforward: In a non-interacting electron system, the
many-body state can be expressed as a Slater determinant in the zero-temperature
limit, which can be fully determined by the first-order correlation function. By
calculating the correlation function based on the scattering matrix theory of
quantum transport, one can reconstruct the Slater determinant from the
scattering matrix.

\begin{figure}
  \includegraphics[width=7.0cm]{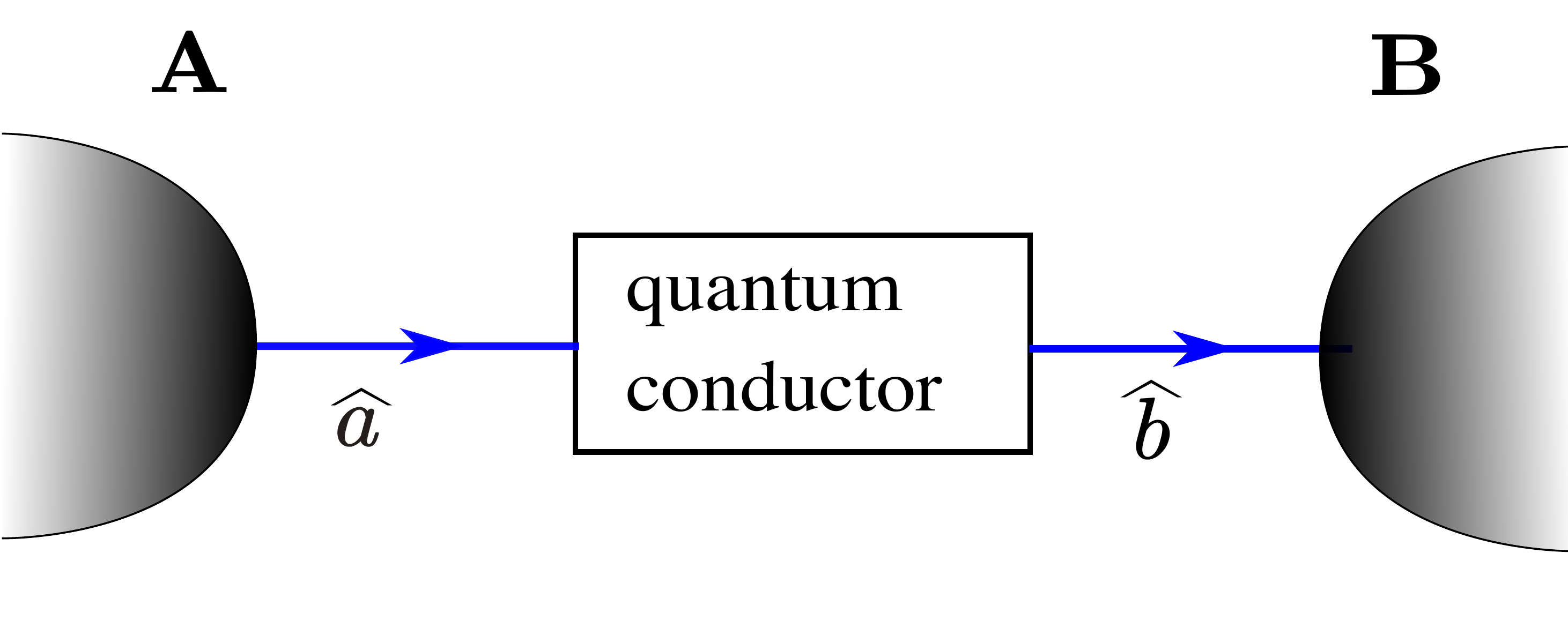}
  \caption{ (a) A single-channel quantum conductor connected to two reservoirs
    $\mathbf{A}$ and $\mathbf{B}$.}
  \label{figa1}
\end{figure}

Such reconstruction can be demonstrated more clearly in a single-channel quantum
conductor at zero temperature, as illustrated in Fig.~\ref{figa1}. The incoming
electrons are injected from the reservoir $\mathbf{A}$ into the conductor, while
the outgoing electrons from the conductor are fed into the reservoir
$\mathbf{B}$. Without interactions, the quantum transport of electrons in such
system can be generally described by a single-body scattering matrix
$\mathbf{S}$. By introducing annihilation operators $\hat{a}(E)$ and
$\hat{b}(E)$ for the incoming and outgoing electrons in the energy domain, one
has
\begin{equation}
  \hat{b}(E) = \sum_{E'} \mathbf{S}(E, E') \hat{a}(E'),
  \label{a1:eq1}             
\end{equation}
with $\mathbf{S}(E, E')$ representing the matrix element of the scattering
matrix $\mathbf{S}$ in the energy domain.

It is convenient to introduce the polar decomposition of the scattering matrix,
which has the form
\begin{eqnarray}
  \mathbf{S}(E, E') & = & \sum_j \left[ \begin{tabular}{cc}
                                          $\psi^e_j(E)$, & $\psi^h_j(E)$\\
                                        \end{tabular}\right] \nonumber\\
                    &\times& \left[\begin{tabular}{cc}
                                     $\sqrt{ 1 - p_j }$ & $i \sqrt{p_j}$\\
                                     $i \sqrt{p_j}$ & $\sqrt{ 1 - p_j}$\\
                                   \end{tabular}\right] \left[\begin{tabular}{c}
                                                                $\varphi^e_j(E')$\\
                                                                $\varphi^h_j(E')$\\
                                                              \end{tabular}\right]^{\ast},
  \label{a1:eq1-1}
\end{eqnarray}
where $\psi^e_j(E)$, $\varphi^e_j(E')$ are nonzero for $E > 0$, while
$\psi^h_j(E)$, $\varphi^h_j(E')$ are nonzero for $E \le 0$. These functions form
orthonormal basis in the energy domain:
\begin{equation}
  \int \frac{dE}{2\pi\hbar} \langle \alpha'_{j'} | E \rangle \langle E |
  \alpha_j \rangle = \delta_{j, j'} \delta_{\alpha, \alpha'}, 
  \label{a1:eq1-2}           
\end{equation}
with $\delta_{j,j'}$ being the Kronecker delta. Note that we have introduce the
Dirac notation $\langle E | \alpha_k \rangle = \psi^{\alpha}_j(E)$, with
$\alpha = e, h$.

Now let us turn to discuss the many-body state of the electrons in such
system. For the incoming electrons, the many-body state
$|\mathbf{\Psi_A}\rangle$ is just a Fermi sea $|\mathbf{F}\rangle$, whose Fermi
level $E_F$ is decided by the reservoir $\mathbf{A}$. Assuming $E_F=0$, it can
be expressed as
\begin{equation}
  |\mathbf{\Psi_A}\rangle = |\mathbf{F}\rangle = \prod_{\epsilon \le 0} \hat{a}^{\dagger}(\epsilon) |\mathbf{Vac}\rangle,
  \label{a1:eq2}
\end{equation}
with $|\mathbf{Vac}\rangle$ being the vacuum state.

Alternatively, one can also describe the many-body state
$|\mathbf{\Psi_A}\rangle$ by using the corresponding first-order correlation
function, which has the form in the energy domain
\begin{equation}
  i G^{<}_A(E, E') = \langle \mathbf{F} | \hat{a}^{\dagger}(E') \hat{a}(E) | \mathbf{F} \rangle,
  \label{a1:eq2-1}
\end{equation}
or equivalently, 
\begin{equation}
  i G^{>}_A(E, E') = \langle \mathbf{F} | \hat{a}(E) \hat{a}^{\dagger}(E') | \mathbf{F} \rangle,
  \label{a1:eq2-1a}
\end{equation}
which related to $G^{<}_A(E, E')$ as $G^{<}_A(E, E') + G^{>}_A(E, E') = i\delta(E-E')$.

By substituting Eq.~\eqref{a1:eq2} into Eqs.~\eqref{a1:eq2-1}
and~\eqref{a1:eq2-1a}, the correlation function $G^{\gtrless}_A$ can be
expressed by using the single-body states $| \epsilon \rangle$ of electrons as
\begin{eqnarray}
  i G^{<}_A(E, E') & = & \sum_{\epsilon \le 0} \langle E | \epsilon \rangle
                       \langle \epsilon | E' \rangle, \nonumber\\
  i G^{>}_A(E, E') & = & \sum_{\epsilon > 0} \langle E | \epsilon \rangle
                       \langle \epsilon | E' \rangle, \label{a1:eq2-2a}
\end{eqnarray}
with $\langle E | \epsilon \rangle = \delta(E-\epsilon) \Delta E$. Here
$\Delta E$ is the mesh size in the energy domain. The limit $\Delta E \to 0$
should be taken in the end of the calculation.

The many-body state $|\mathbf{\Psi_B}\rangle$ of the outgoing electrons is
usually not given explicitly in the scattering matrix theory. Instead, it is
described by the first-order correlation function as:
\begin{align}
  \label{a1:eq3}  
  &G^{<}_B(E, E')= \langle \mathbf{F} | \hat{b}^{\dagger}(E') \hat{b}(E) |
    \mathbf{F} \rangle \nonumber\\
  &= \sum_{E_1, E'_1} \langle \mathbf{F} | [\mathbf{S}(E', E'_1)]^{\ast}
    \mathbf{S}(E, E_1) \hat{a}^{\dagger}(E'_1)
    \hat{a}(E_1) | \mathbf{F} \rangle.
\end{align}

To find the explicit form of the many-body state $|\mathbf{\Psi_B}\rangle$, we
write $G^{\gtrless}_B(E, E')$ in a form analogous to
Eqs.~\eqref{a1:eq2-2a}. This can be done by using the polar decomposition of the
scattering matrix given in Eq.~\eqref{a1:eq1-1}, which gives
\begin{equation}
  G^{\gtrless}_B(E, E') = \sum_j \langle E | \gamma^{\gtrless}_j \rangle \langle \gamma^{\gtrless}_j | E' \rangle,
  \label{a1:eq3-1}
\end{equation}
with
\begin{eqnarray}
  | \gamma^{<}_j \rangle & = & i \sqrt{p_j} | e_j \rangle + \sqrt{1-p_j} | h_j
                               \rangle, \nonumber\\
  | \gamma^{>}_j \rangle & = & i \sqrt{p_j} | h_j \rangle + \sqrt{1-p_j} | e_j
                               \rangle.
                               \label{a1:eq3-2}                               
\end{eqnarray}

This indicates that the many-body state $|\mathbf{\Psi_B}\rangle$ can be
expressed in a BCS-like form, corresponding to a neutral cloud of $e$h
pairs. The quantum state of the $e$h pair can be described by the excitation
probability $p_k$ and the single-body state $|e_k\rangle$ [$|h_k\rangle$] of the
electron [hole] components. They can be obtained by solving the polar
decomposition of the scattering matrix. In our previous works
\cite{yin-2019-quasip-states, yue-2019-normal-anomal}, we have studied the
quantum state of $e$h pairs by using such decomposition.

It is possible for the polar decomposition to give solutions corresponding to
either $| e_j \rangle = 0$ or $| h_j \rangle = 0$, when $p_j = 1.0$. This
indicates that there also exist unpaired electrons or holes, which are just
quasiparticles carrying negative or positive charges. Moreover, additional
normalization factors can also emerge, representing the injection probability of
the corresponding quasiparticles. This is the case we have encountered in this
paper, when the corresponding first-order correlation function can be given as
\begin{equation}
  G^{\gtrless}_B(E, E') = \sum_j q_j \langle E | \gamma^{\gtrless}_j \rangle \langle \gamma^{\gtrless}_j | E' \rangle.
  \label{a1:eq4}
\end{equation}
By taking these ingredients into consideration, one can express the many-body
state as given in Eqs.~\eqref{s1:eq1},~\eqref{s1:eq2} and~\eqref{s1:eq3}. The
correlation function in the time domain can be obtained from the transform
\begin{equation}
  G^{\gtrless}(t, t') = \int \frac{dEdE'}{(2\pi\hbar)^2} e^{-iEt/\hbar+iE't'/\hbar} G^{\gtrless}(E, E').
  \label{a1:eq5}
\end{equation}
Note that in the main text, we have expressed the first-order correlation
function $G^{<}_B(t, t') = G(t, t')$ in the time domain [see
Eq.~\eqref{s2:eq12}].

For the system we considered here, the scattering matrix has a simple structure
in the time domain [see Eq.~\eqref{s2:eq1}]. By introducing the wave packet
functions
\begin{align}
  \varphi^{e/h}_{kl}(t) &= \int \frac{dE}{2\pi\hbar\sqrt{q_k}} e^{-iE(t-lT/q_k)/\hbar}
                       \varphi^{e/h}_k(E), \nonumber\\
  \psi^{e/h}_{kl}(t) &= \int \frac{dE}{2\pi\hbar\sqrt{q_k}} e^{-iE(t-lT/q_k)/\hbar}
                       \psi^{e/h}_k(E),
  \label{a1a:eq2}
\end{align}
with $q_k$ being the normalization factor, the polar decomposition
[Eq.~\eqref{a1:eq1-1}] can be obtained by solving the equations
\begin{equation}
  e^{-i \phi(t)} \left[\begin{tabular}{c}
                         $\varphi^e_{kl}(t)$\\
                         $\varphi^h_{kl}(t)$\\
                       \end{tabular}\right] = \sqrt{1-p_k} \left[\begin{tabular}{c}
                                                                   $\psi^e_{kl}(t)$\\
                                                                   $\psi^h_{kl}(t)$\\
                                                                 \end{tabular}\right]
                                                               + i \sqrt{p_k} \left[\begin{tabular}{c}
                                                                                      $\psi^h_{kl}(t)$\\
                                                                                      $\psi^e_{kl}(t)$\\
                                                                                    \end{tabular}\right],
                                                                                  \label{a1a:eq1}
\end{equation}
where we have chosen the compound index $j=[k, l]$. Here $\phi(t)$ is the
forward scattering phase, which can be written as
$\phi(t) = \frac{e}{\hbar} \int^t V(\tau) d\tau$.

Although $V(t)$ is periodic, the forward scattering phase $\phi(t)$ is non-periodic. In fact, it is possible to extract
the periodic part from $\phi(t)$ by introducing the relation
\begin{equation}
  \varphi = N_T + \frac{\omega_T}{\Omega},
  \label{a1a:eq3}
\end{equation}
with $N_T$ being integer and $\omega_T$ being real number, which satisfies $\omega_T \in [0, \Omega]$. By using $N_T$
and $\omega_T$, one can express $\phi(t)$ as
\begin{equation}
  \phi(t) = \phi_n(t) + [\omega_T + (N_T - n)\Omega]t.
  \label{a1a:eq3-1}
\end{equation}
where $\phi_n(t)$ represents the periodic part of the forward scattering phase, with $n$ being integer.

Hence the integer $n$ offers a natural index for the solutions. For a given $n$, Eqs.~\eqref{a1a:eq2}
and~\eqref{a1a:eq1} can be reduced to a singular value problem, whose solutions can be labelled by another integer
$m$. That is why we choose the index $k=[n, m]$ in the main text. Note that for a given flux $\varphi=Q/e$, we find that
only the solutions related to $n=N_T$ and $n=N_T-1$ are relevant. The corresponding $\psi^{e/h}_k(t)$ can be expressed
by the ansatz:
\begin{equation}
  \psi^{e/h}_{kl}(t) = U^{e/h}_k(t) \int \frac{d\omega}{2\pi\sqrt{q_k}} F^{Q}_k(\omega)
  e^{-i\omega (t-T/q_k)},
  \label{a1a:eq4}
\end{equation}
where $F^Q_k(\omega)$ is the real function given in Eq.~\eqref{s2:eq9}. The function $U^{e/h}_k(t)$ is periodic in the
time domain, which usually has to be obtained numerically from the singular value problem.

\section{Current and shot noise}
\label{app2}

Various observable quantities can be calculated directly from the expression of the many-body state
$|\mathbf{\Psi_B}\rangle$ given in Appendix~\ref{app1}. The current carried by the train of wave packets can be given as
\begin{align}
  &I(t) = e\langle \mathbf{\Psi_{\rm train}}| \hat{a}^{\dagger}(t) \hat{a}(t)
    | \mathbf{\Psi_{\rm train}} \rangle \nonumber\\
  & = \sum_k\sum_{l=0, \pm 1, \pm 2, ...} I^c_{kl}(t) + \sum_k\sum_{l=0, \pm 1, \pm 2, ...} I^{eh}_{kl}(t),
    \label{a2:eq1}             
\end{align}
where $I^c_k(t)$ represents the contribution from the charged quasiparticles, which has the form
\begin{equation}
  I^c_{kl}(t) = e q_k|\psi^c_{kl}(t)|^2.
  \label{a2:eq1-1}
\end{equation}
In contrast, $I^{eh}_{kl}(t)$ represents the contribution from the $e$h pair, which can be written as
\begin{align}
  I^{eh}_{kl}(t) &= e q_k p_k [ |\psi^e_{kl}(t)|^2 - |\psi^h_{kl}(t)|^2 ] \nonumber\\
                 &\mbox{}+ 2 e q_k \sqrt{p_k(1-p_k)}
                   \operatorname{Im}\{ \psi^h_{kl}(t) [\psi^e_{kl}(t)]^{\dagger}  \}.  
    \label{a2:eq1-2}
\end{align}
Note that for the electron source we considered here, one always has $I(t) = (e^2/h) V(t)$.

When the train of wave packets is partitioned at a localized scatter with transmission probability $D$, both the
quasiparticles and $e$h pairs can contribute to the shot noise. The time-dependent shot noise can be expressed as
\begin{align}
  \hspace{-0.0cm}S_{sn}(t, t') &= e^2D(1-D) \frac{i}{2\pi} \frac{1}{t-t' + i \eta}\nonumber\\
                               &\hspace{-1cm}\times \sum_k q_k \sum_{l=0, \pm 1, \pm 2, ...} \Big\{ \psi^c_{kl}(t) [\psi^c_k(t')]^\dagger \nonumber\\
                               &\hspace{-0.75cm}\mbox{}+\Big[ \sqrt{p_k} \psi^e_{kl}(t) - i \sqrt{1-p_k}
                                 \psi^h_{kl}(t) \Big] \nonumber\\
                               &\hspace{-0.5cm}\times \Big[ \sqrt{p_k} \psi^e_{kl}(t') - i
                                 \sqrt{1-p_k} \psi^h_{kl}(t') \Big]^\dagger \nonumber\\
                               &\hspace{-0.75cm}\mbox{}+ \Big[ \sqrt{p_k} \psi^h_{kl}(t') - i
                                 \sqrt{1-p_k} \psi^e_{kl}(t') \Big] \nonumber\\
                               &\hspace{-0.5cm}\times \Big[ \sqrt{p_k} \psi^h_{kl}(t) - i
                                 \sqrt{1-p_k} \psi^e_{kl}(t) \Big]^\dagger \Big\}.
                                 \label{a2:eq2} 
\end{align}

The shot noise in the dc limit can be obtained as
\begin{eqnarray}
  \overline{S_{sn}(t, t')} & = & \int^{+\infty}_{-\infty} \int^{+\infty}_{-\infty} S_{sn}(t, t') dt dt' \nonumber\\
                           & = & \sum_{l=0, \pm 1, \pm 2, ...} [S_{c}(l) + S_{ex}(l)],
                                 \label{a2:eq3}
\end{eqnarray}
where $S_{c}(l)$ and $S_{ex}(l)$ representing the shot noise attributed to the charged quasiparticles and $e$h pairs in
the $l$-th wave packet. We find that both of them are independent on the index $l$, which can be written as
\begin{align}
  S_c &= S_0 \sum_k q_k, \nonumber\\
  S_{ex} &= 2 S_0 \sum_k q_k p_k .
  \label{a2:eq4}
\end{align}
with $S_0 = 2 \frac{e^2}{h} D(1-D)\hbar\Omega$ being the typical scale of the shot noise.

\section{Waiting time distribution}
\label{app3}

To obtain the information of the many-body state in short time scales, we study the waiting time distribution between
electrons above the Fermi sea. This can be obtained from the corresponding time-dependent full counting statistics (FCS)
\cite{dasenbrook16_quant_theor_elect_waitin_time_clock}. As far as the electrons above the Fermi sea are concerned, the
characteristic function of the corresponding FCS in the time interval $[t_s, t_e]$ can be given as
\begin{equation}
  \chi(\lambda; t_s, t_e) = \langle \mathbf{\Psi_B} | e^{i \lambda
    \hat{N}_{se}} |\mathbf{\Psi_B}\rangle,
  \label{a3:eq1}             
\end{equation}
where the operator $\hat{N}_{se}$ counts the number of electrons above the Fermi sea injected in the time interval
$[t_s, t_e]$. It can be given as
\begin{equation}
  \hat{N}_{se} = \int^{t_e}_{t_s} dt \hat{a}^{\dagger}_p(t) \hat{a}_p(t),
  \label{a3:eq2}             
\end{equation}
with $\hat{a}_p(t) = \int^{+\infty}_0 e^{-iEt/\hbar} \hat{a}(E) dE/(2\pi\hbar)$.

By using the expression of $|\mathbf{\Psi_B}\rangle$, the characteristic function can be expressed as
\begin{equation}
  \chi(\lambda; t_s, t_e) = \det[ \hat{1} + (e^{i\lambda} - 1) \hat{\Lambda}_{se} \hat{G}^{<}_B ].
  \label{a3:eq3} 
\end{equation}
with $\hat{G}^{<}_B = \sum_j q_j | \gamma^{<}_j \rangle \langle \gamma^{<}_j|$ being the operator corresponding to the
first-order correlation function. Here we have introduced a single-body operator
$\hat{\Lambda}_{se} = \int^{t_e}_{t_s} dt |t_p\rangle \langle t_p|$ corresponding to $\hat{N}_{se}$, with
$|t_p\rangle = \int^{+\infty}_0 e^{iEt/\hbar}|E\rangle dE/(2\pi\hbar)$. This allows us to calculate the FCS by using the
information of the single-body states of the charged quasiparticles and $e$h pairs. By taking the limit
$\lambda \to i \infty$, the above equation can be reduced to the idle time probabilities of electrons above the Fermi
sea [see also Eq.~\eqref{s3:eq7}]
\begin{equation}
  \Pi(t_s, t_e) = \det[ \hat{1} - \hat{Q}_{se} ].
  \label{a3:eq4} 
\end{equation}
with $\hat{Q}_{se} = \hat{\Lambda}_{se} \hat{G}^{<}_B$.

%

\end{document}